\newcommand{\tsecompldate}{6th February 2007}
\documentclass[12pt]{article}

\usepackage{amsmath,amssymb,amscd}

\usepackage{graphicx}

\newcommand{\vol}[1]{\textbf{#1}}

\newcommand{\tpaptitle}[1]{``#1'',} 
\newcommand{\tpretitle}[1]{``#1'',} 
\newcommand{\tarttitle}[1]{``#1'',} 
\newcommand{\tbktitle}[1]{``#1''}     
\newcommand{\tISBN}[1]{#1} 

\newcommand{\tref}[1]{(\ref{#1})}

\newcommand{\tnotpre}[1]{#1} 
\newcommand{\tversion}{Version 2}

\newcommand{\tpre}[1]{#1} 
\newcommand{\tprenote}[1]{} 

\newcommand{\tnote}[1]{} 
\newcommand{\tcomment}[1]{} 
\newcommand{\tcommentx}[1]{} 

\newcommand{\href}[2]{#2}
\newcommand{\eprint}[1]{\texttt{#1}}

\newcommand{\tsedevelop}[1]{{}} 

\def\tsetrue{T} \def\tsefalse{F} 

\let\tsepaper=\tsefalse   
\let\tseprivateon=\tsefalse   
\let\tseletter=\tsefalse  
\let\tsedevon=\tsefalse    
\let\tsehypertexton=\tsetrue 
\if\tsetrue\tsepaper  \else
 \fi

\if\tsetrue\tsedevon 
\usepackage{showkeys} 
\renewcommand{\tsedevelop}[1]{{#1}}
\fi

\if\tsefalse\tsepaper

\fi

\if\tsetrue\tseletter   \fi

\renewcommand{\vol}[1]{{\bf #1}}

\renewcommand{\tpaptitle}[1]{\emph{#1},}

\renewcommand{\tarttitle}[1]{\emph{#1},}

\renewcommand{\tbktitle}[1]{``#1''}
\renewcommand{\tISBN}[1]{}

\renewcommand{\tpretitle}[1]{{\em #1},}

\if\tsefalse\tsehypertexton \renewcommand{\href}[2]{{#2}{}}
\renewcommand{\eprint}[1]{\href{http://xxx.soton.ac.uk/abs/#1}{{\tt #1}}}
\fi

\if\tsefalse\tsepaper \renewcommand{\tnotpre}[1]{}
\renewcommand{\tpre}[1]{#1}
\renewcommand{\tprenote}[1]{\footnote{#1}}
\renewcommand{\tversion}{Preprint version (some additional comments and material)} \fi

\if\tsetrue\tseprivateon 
\renewcommand{\tnote}[1]{\footnote{({\cal T}) #1}}
\renewcommand{\tcomment}[1]{#1}
\renewcommand{\tversion}{Private version} \fi

\newcommand{\half}{\frac{1}{2}}
\newcommand{\bea}{\begin{eqnarray}}
\newcommand{\eea}{\end{eqnarray}}
\newcommand{\beq}{\begin{equation}}
\newcommand{\eeq}{\end{equation}}
\newcommand{\nnel}{\nonumber \\ {}}

\newcommand{\ra}{\rightarrow}

\newcommand{\dprime}{{\prime\prime}}

\newcommand{\Eopp}{\frac{E}{p_p}}
\newcommand{\Etilde}{\frac{p_r}{p_p}E}
\newcommand{\Ktilde}{\frac{p_r}{p_p}\kav}

\newcommand{\psharp}{p_\sharp}
\newcommand{\pstar}{p_\ast}

\newcommand{\taverage}[1]{\langle #1 \rangle}
\newcommand{\kav}{\taverage{k}}
\newcommand{\efunc}{\omega}

\begin{document}

\renewcommand{\thefootnote}{\fnsymbol{footnote}}

 \tpre{
 \begin{flushright}
 \texttt{Imperial/TP/06/TSE/5} \\
 \eprint{cond-mat/0612214} \\
 \tsecompldate
 \\ Published version: Phys.Rev.E.\vol{75} (2007) 056101
 \tpre{\\ \tversion  }
 \tsedevelop{\\(\texttt{rwlong2arXiv.tex}  LaTeX-ed on \today ) }
 \end{flushright}
 \vspace*{0.5cm}
 }

\begin{center}
{\Large\textbf{Exact Solution for the Time Evolution of Network
Rewiring Models}}\tnote{tnotes such as this not present in final
version} \\[0.5cm]
\tpre{\vspace*{1cm} }
 {\large T.S.\ Evans\footnote{\href{http://www.imperial.ac.uk/people/t.evans}{\texttt{http://www.imperial.ac.uk/people/t.evans}} }},
 {\large A.D.K.\ Plato}
 \\[0.5cm]
 \tpre{\vspace*{1cm}}
 \href{http://www.imperial.ac.uk/research/theory}{Theoretical Physics},
 Blackett Laboratory, Imperial College London,\\
 South Kensington campus, London, SW7 2AZ,  U.K.
\end{center}


\begin{abstract}
We consider the rewiring of a bipartite graph using a mixture of
random and preferential attachment.  The full mean field equations
for the degree distribution and its generating function are given.
The exact solution of these equations for all finite parameter
values at any time is found in terms of standard functions. It is
demonstrated that these solutions are an excellent fit to numerical
simulations of the model.  We discuss the relationship between our
model and several others in the literature including examples of
Urn, Backgammon, and Balls-in-Boxes models, the Watts and Strogatz
rewiring problem and some models of zero range processes.  Our model
is also equivalent to those used in various applications including
cultural transmission, family name and gene frequencies, glasses,
and wealth distributions. Finally some Voter models and an example
of a Minority game also show features described by our model.
\end{abstract}

\renewcommand{\thefootnote}{\arabic{footnote}}
\setcounter{footnote}{0}

\section{Introduction}

One of the the most important classes of complex network models are
those with a constant number of edges which evolve by rewiring those
edges.  The classic example of Watts and Strogatz \cite{WS98} is of
this type and such models are often studied in their own right
\cite{BCK01,DM03,DMS03,PLY05,XZW05,OTH05}. Network rewiring is also
related to to some multi-Urn models \cite{GBM95,GL02,OYT05,OYT06}
which include what are termed Backgammon or Balls-in-Boxes models
\cite{BBBJ00} used for glasses \cite{Ritort95,BBJ97}, simplicial
gravity \cite{BBJ99} and wealth distributions \cite{BDJKNPZ02}.
Models of the zero range process \cite{EvansMR00,EH05,PM05} are also
closely linked. Since most practical systems cannot grow
indefinitely networks of constant size have many applications: the
transmission of cultural artifacts such as pottery designs, dog
breed and baby name popularity
\cite{Neiman95,BS03,HB03,HBH04,BHS04,BS05},  the distribution of
family names in constant populations \cite{ZM01}, the diversity of
genes \cite{KC64,CK70}.  Aspects of the Voter model
\cite{Liggett85,SR05}, as used to describe the competition between
languages \cite{SCEM06}, and the popularity of minority game
strategies \cite{ATBK04} may also be cast in terms of network
rewiring.

Analytic results for network models are limited. A typical approach
starts from the master equations for the evolution of the degree
distribution, these are given in a mean field approximation in which
the quantities are the average values of many possible realisations.
Luckily in most models, and even in some real-world applications,
the results from mean field equations often agree extremely well
with numerical simulations of the model.

Despite the simplifications brought by the mean field approximation,
the equations remain difficult to solve and it is normal to study
the large graph and long time limit e.g.\ see \cite{DM01,KR01}. For
instance the finite size/time corrections to growing graphs using
linear degree attachment probabilities are complicated and known
only as an asymptotic expansion, for example see \cite{KR02,ES05}.
In fact one of the most tractable examples remains the
Erd\H{o}s-R\'eyni random graph which can be seen as the long time
limit of the Watts and Strogatz rewiring model \cite{WS98}.

What we show in this paper is that the mean field equations for the
degree distribution of non-growing rewiring models with linear
rewiring probabilities can be solved \emph{exactly} for \emph{any}
time. This goes beyond the results found in the literature
(typically exact only for infinitely large systems in equilibrium)
and extends the initial work on exact results in such models at
equilibrium \cite{Evans06} and the preliminary non-equilibrium
results of \cite{EP06ECCS}.

We start by setting up the model and the mean field master equations
for the degree distribution in the next Section.  We solve these in
terms of the generating function in Section \ref{sec:genfunc} and
from this we consider the degree distribution in Section \ref{secDD}
and then its moments in Section \ref{secmoments}.  All this is done
in terms of our simple network rewiring model but in
Section~\ref{discussion} we consider the relation between this model
and a variety of other abstract models (with and without explicit
networks) and to various real world examples.  Finally we summarise
our conclusions and add some observations on how such preferential
attachment may arise naturally and the scaling properties of our
model.

\section{The Model}\label{sec:model}

\begin{figure}[hbt]
{\centerline{\includegraphics[width=8cm]{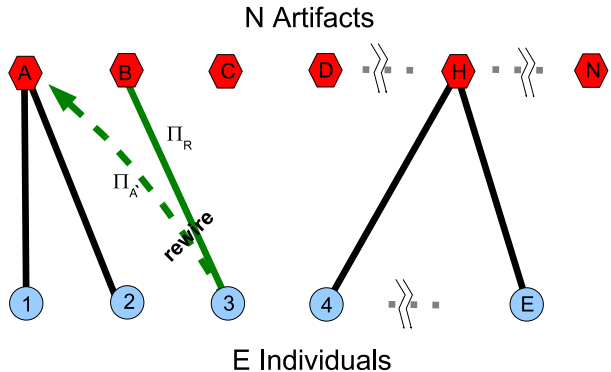}}}
\caption{The bipartite graph has $E$ `individual' vertices, each
with one edge. The other end of the edge is connected to one of $N$
`artifact' vertices. If the degree of an artifact vertex is $k$ then
this artifact has been `chosen' by $k$ distinct individuals. At each
time step a single rewiring of the artifact end of one edge occurs.
An individual is chosen (number 3 here) with probability $\Pi_R$
which gives us the departure artifact (here B).  At the same time
the arrival artifact is chosen with probability $\Pi_A$ (here
labelled A). After both choices have been made the rewiring is
performed (here individual 3 switches its edge from artifact B to
A).}
 \label{fCopyModel2}
\end{figure}

We will focus on a generic rewiring problem, which we shall
describe in terms of a bipartite graph of $E$ `individual'
vertices, each having one edge fixed to any one of $N$ `artifact'
vertices, as shown in Fig.~\ref{fCopyModel2}.  Our naming of the
vertices reflects our previous work and one possible application
(cultural transmission) but apart from the names we will keep our
presentation abstract until Section~\ref{discussion}.

Each individual vertex is always connected to exactly one
edge\tnote{This is not important for our results but is useful when
making contact with other models in the literature.} while the other
end of each edge is connected to any artifact. The network changes
by rewiring the artifact end of these edges and we will focus on the
degree distribution of the artifact vertices at any one time,
$n(k,t)$ and its probability distribution $p(k,t)=n(k,t)/N$ where
$k$ is the degree of an artifact vertex.

To make progress we make further simplifying assumptions.  First
we will assume that the population of individuals is absolutely
constant so $E$ is fixed and finite.  Almost all other comparable
work uses a large $E$ approximation. We will also assume that the
artifact choices available are fixed to be $N$ so the average
degree of an artifact vertex is $\kav=E/N$.  An important limit is
where we take $N$ to infinity so $\kav \rightarrow0$.

We will then assume that at each time step one edge is
rewired\footnote{Alternatively the order in which each individual
changes their choice could be made in a more systematic way, either
in a fixed order or a random order changed once each individual has
been rewired once. They could even be changes at the same time, the
model used in \cite{BS03,HB03,HBH04,BHS04}. We have tried these
variations numerically and they appear to make little difference to
the equilibrium results.}. Continuous time evolution is considered
in Section \ref{ssgenmodel}. At each time step\footnote{The physical
time $\mathcal{T}$ of each event $t \in \mathbb{Z}$ should be
monotonically increasing $\mathcal{T}(t+1) > \mathcal{T}(t)$ but
otherwise it can be arbitrary. This relationship should be derived
from the actual frequency of changes in the problem of interest.} we
first make two choices and only then do we change the network.

First an individual is chosen in some stochastic manner. This
individual is attached by one edge to an artifact, the departure
artifact.  It is the artifact end of this edge which is to be
changed. Thus we are effectively removing an edge from the departure
artifact chosen with probability\footnote{Alternatively we also have
models where the process chooses an edge or artifact directly, with
the probability $\Pi_R$.  The distinction is immaterial for the
degree distribution of the artifacts so we shall take these
alternatives for granted.} $\Pi_R$. The edge chosen is going to be
rewired and attached to another artifact vertex, the arrival
artifact, picked with probability $\Pi_A$. Thus the master equation
for the degree distribution in the mean field approximation is
\begin{eqnarray}
\lefteqn{n(k,t+1) - n(k,t)}
 \nnel
 &=&   n(k+1,t) \Pi_R(k+1,t) \left( 1- \Pi_A(k+1,t) \right)
 \nnel
 &&
     - n(k,t)   \Pi_R(k,t)   \left( 1- \Pi_A(k,t)   \right)
     - n(k,t)   \Pi_A(k,t)   \left( 1- \Pi_R(k,t)   \right)
 \nnel
 &&
     + n(k-1,t) \Pi_A(k-1,t) \left( 1- \Pi_R(k-1,t) \right)
   ,
   \qquad (E \geq k \geq 0) \, .
   \label{neqngen}
\end{eqnarray}
For notational simplicity we choose to set
$n(k)=\Pi_R(k)=\Pi_A(k)=0$ for the unphysical values $k=-1$ and
$k=(E+1)$.  In this way the equation gives the correct behaviour at
the physical boundary values of $k=0$ and $k=E$.

Note that there is a chance $(\Pi_R\Pi_A)$ that we will choose the
same artifact vertex for both attachment and removal.  As this
produces no change in the network we must ensure that such events do
not contribute to changes in the degree distribution. This is the
role of the factors of $(1-\Pi)$.  Such terms are not normally found
in the master equations for network rewiring
\cite{DM03,PLY05,XZW05,OTH05,GBM95,GL02,OYT06}.\tnote{Equations (4)
in \cite{XZW05}, (3) in \cite{PLY05}, (2) in \cite{OYT06}, and (6)
in \cite{OTH05}. Equation (2.13) \cite{GL02} and there is no
dependence in the rate $W$ on the number of boxes with degree $k$
when $k=l$. If one tries to ensure true Urn model behaviour, no
tadpoles in the box connection network, no same box choices, then
the $\phi(k)$ function in section 6.2.4 has to depend of the
departure degree through the normalisation as well as on the arrival
degree. (4.6) in \cite{DM03} but there is correct discussion of this
being for safe graphs in large graph limit.} It is \emph{crucial}
that we do this otherwise we will not have the correct behaviour at
the boundaries $k=0$ and $k=E$.

Such $(1-\Pi)$ corrections will often be negligible especially for
large $E$ systems where the probabilities $\Pi(k)$ for any
individual value of degree $k$ may be tiny\footnote{These are the
``safe situations'' of large graphs discussed in Chapter 4 of
\cite{DM03}.}. However there are important configurations in this
model and in related models where even for large systems in
equilibrium $\Pi(k)$ are not small for some values of $k$. This will
be discussed in Section \ref{ssgenmodel} after we have obtained the
explicit solution.

The master equation \tref{neqngen} is a mean-field approximation for
the evolution through our stochastic dynamics of the average value
of the function $n(k)$.   The errors come if
\beq
 \langle n(k,t) f(k,t) \rangle - \langle n(k,t) \rangle \langle f(k,t) \rangle
 \neq 0
 \label{nonfactor}
\eeq
where $f$ are various combinations of $\Pi_R$ and $\Pi_A$.  For
instance if we have a model with attachment or removal probabilities
of the form $(k^\beta/z_\beta)$ then the problem lies with the
normalisation as in general
\beq
 \langle n(k,t) \frac{k^\beta}{z_\beta(t)} \rangle
 \neq
 k^\beta \frac{\langle n(k,t) \rangle }{\langle z_\beta(t) \rangle}
 \; .
 \label{nonlinprob}
\eeq
In many practical cases the fluctuations are small and the
corrections to the mean-field results are often found to be small.
For this reason the equations can be a good approximation even if
the number of vertices or edges fluctuate provided their average
values are constant and the variations are small.

However, there are two special cases where equality holds in
\tref{nonlinprob} implying that the mean field approximation is
exact, namely when $\beta=0$ or $\beta=1$. Only in these cases are
the normalisations of probabilities constants of the motion, $N$ and
$E$ respectively. The most general choice for $\Pi_R$ and $\Pi_A$
satisfying these criteria is therefore
\beq
 \Pi_R = \frac{k}{E}, \qquad
 \Pi_A = p_r\frac{1}{N} + p_p\frac{k}{E},
   \qquad p_p+p_r=1 \qquad (E \geq k \geq 0) \; .
 \label{PiRPiAsimple}
\eeq
This form for $\Pi_A$ means that an edge can be reattached in two
ways. With probability $p_p$, \emph{preferential attachment} is used
and the artifacts are chosen with a likelihood proportional to their
degree. Alternatively with probability $p_r$ a random\footnote{In
this paper `random' without further qualification indicates that a
uniform distribution is used to draw from the set implicit from the
context.} artifact is chosen.  Choosing a random edge for rewiring
corresponds to the use of `preferential removal' alone.

There are other good reasons for choosing these forms for the
probability apart from the fact the master equation is then exact.
Mathematically, these simple forms enable us to find a complete
non-equilibrium solution.  In terms of practical applications, one
may understand these special forms as emerging naturally from a
random walk process \cite{ES05,SK04}.  We will also note the scaling
properties of their solutions in Section \ref{sec:sumconcl}.  We
will limit our analysis to the case \tref{PiRPiAsimple} which means
we will present exact results for the ensemble average of various
quantities at any time and for any values of our parameters.

\section{The Generating Function}\label{sec:genfunc}

A useful way to investigate the degree distribution $n(k,t)$ is to
encode it with a generating function $G(z,t)$
\beq
 G(z,t) := \sum_{k=0}^{E} z^k n(k,t) \, .
 \label{Gktdef}
\eeq
Below we will exploit the fact that $G$ is always a polynomial in
$z$ of order no greater than $E$.  The mean field equations
\tref{neqngen} can then be re-written as a differential equation
for the generating function,
\bea
 \lefteqn{\frac{b(1+a-c) }{(1-z)}    \left[  G(z,t+1)-G(z,t)\right]
 }
 \nnel
 &=&
  z(1-z)G^{\dprime}(z,t)
 +[c-(a+b+1)z]G^{\prime}(z,t)
 -ab G(z,t) \;,
 \label{eq:gendifft}
\eea
where the differentials $G^{\dprime}$ and $G^\prime$ are with
double and single derivatives with respect to $z$. The constants $a$, $b$ and $c$ are given by,
\beq
 a = \frac{p_r}{p_p}\kav \; , \qquad
 b = -E \; , \qquad
 c = 1+\frac{p_r}{p_p}\kav-\frac{E}{p_p}.
\eeq

The equation for $n(k,t)$ is linear---it is completely equivalent to
a Markov process in an $E+1$-dimensional space in which the vector
$(n(0,t), n(1,t) \ldots,n(E,t))$ lives
\cite{EP06ECCS}.\tnote{Perhaps an appendix with the Markov stuff in
it? Poincar\'e-Perron/Perron-Frobenius? Theorem? Is this by virtue
of convergence and solutions being polynomial? should check that
$a=1$ does not exclude Poincar\'e-Perron.} Therefore we can define
$E+1$ eigenvectors $\efunc^{(m)}(k)$ associated with an eigenvalue
$\lambda_m$ ($m=0,1,2,\ldots E)$, which we order such that
$\lambda_m \geq \lambda_{m+1}$.  Furthermore, the properties of the
Markov process guarantee that $1 \geq |\lambda_m|$ with at least
$\lambda_0=1$.

We can now break the generating function into $E+1$ components with
the time dependence factorised
\begin{equation}
 G(z,t) = \sum_{m=0}^{E} c_m (\lambda_m)^t G^{(m)}(z) \; ,
 \qquad
 G^{(m)}(z) := \sum_{k=0}^{E} z^k \efunc^{(m)}(k)
 \label{Gkmdef}
\end{equation}
where the coefficients $c_m$ depend on the initial conditions
$n(k,t=0)$.  Again the generating functions for the eigenvectors,
$G^{(m)}(z)$, are polynomials of degree no larger than $E$.
Substituting this form into \tref{eq:gendifft} gives a a
time-independent differential equation for $G^{(m)}(z)$, the
generating function of the $m$-th eigenvector:
\bea
 z(1-z) G^{(m)\dprime} (z)
 + [c-(a+b+1)z] G^{(m) \prime}(z) &&
 \nnel
 - \left[ ab-\frac{(\lambda_m -1)}{1-z} b(c-a-1) \right] G^{(m)}(z)
 &=& 0 \; .
 \label{eq:gendiffm}
\eea
This can be solved most easily by writing $G$ as a polynomial in
$(1-z)$.  Having the correct form for the master equation and
therefore the correct behaviour at the boundaries ensures that this
gives a finite order polynomial.\tnote{An argument for the
construction of solutions to this equation is outlined in
Appendix~\ref{solveG}, based on the observation that the generating
functions must be polynomial.} These may be summarised in terms of
the Hypergeometric function $F = {}_2F_1$ to be \footnote{We have
chosen to normalise the eigenfunctions such that
$\efunc^{(m)}(k=0)=1$. The physical normalisation needed in the
problem is contained in the $c_m$ coefficients of
\tref{Gkmdef}.}\tnote{Can we factorise the eigenvalues? I am sure
this must simplify further as the Maple results for the entries of
the eigenvectors $\efunc^{(m)}$ were not so complicated. Is this a
known standard transformation of a hypergeometric function? Expand
binomial as $\Gamma$'s, collect powers of z, do one of the two
sums?}
\bea
 G^{(m)}(z) &=&  (1-z)^m F(a+m,b+m;c;z)
 \\
 &=& (1-z)^m
 \sum_{l=0}^{E-m}\frac{\Gamma(a+m+l)\Gamma(b+m+l)\Gamma(c)}{\Gamma(a+m)\Gamma(b+m) \Gamma(c+l) (l!)} z^l
 \label{Gmresult}
\eea
with corresponding eigenvalues,
\beq
\lambda_{m} =
1 -m(m-1) \frac{p_p}{E^2} - m \frac{p_r}{E},
 \qquad 0 \leq m \leq E \; .
 \label{eq:evalues}
\eeq
An expression for the entries of the eigenvectors $\efunc^{(m)}(k)$
may be derived from the coefficients of $z^k$ in \tref{Gmresult}
which can be given in terms of the Hypergeometric function ${}_3F_2$, though
it is not very
illuminating and merely assists in explicit evaluations:\tnote{TSE not checked this.  Is it OK?  Any further
simplifications?!!!}
\begin{eqnarray}
\efunc^{(m)}(k)&=&(-1)^m \frac{\Gamma(k+1)}{\Gamma(k+1-m)}
\frac{\Gamma(c+k)}{\Gamma(c+k-m)}
\frac{\Gamma(a)}{\Gamma(a+m)} \frac{\Gamma(b)}{\Gamma(b+m)} \times\nonumber\\
&& \qquad \left._3F_2\right. \left.\left(-m,a+k,b+k;k+1-m,k+c-m;1-z
\right)\right|_{z=0} \efunc^{(0)}(k)
 \label{omegamdef}
\end{eqnarray}
 where
\beq
\efunc^{(0)}(k) =
 \frac{\Gamma(a+k)}{\Gamma(a)}
 \frac{\Gamma(b+k)}{\Gamma(b)}
 \frac{\Gamma(c)}{\Gamma(c+k)} \; .
\eeq
In the special case of $p_r=1$, the degree distribution is that of
the Watts and Strogatz model \cite{WS98} (see Section
\ref{discussion} below).  The generating function then reduces
to\tnote{The term to the power $(E-m)$ has the interpretation that
it gives the distribution where one adds $E-m$ edges randomly
i.e.\ with probability $N^{-1}$.}
\beq
G^{(m)} (z) = \frac{1}{(1-N^{-1})^{E-m}}(1-z)^m
((1-N^{-1})+N^{-1}z)^{E-m} \; .
\eeq

More usefully we note for later use that the eigenvalues
satisfy\footnote{The only case of eigenvalue crossing occurs at
$p_r=0$ when there is degeneracy in the two largest eigenvalues
with $1=\lambda_0 = \lambda_1$.}
\bea
1=\lambda_0 > \lambda_1 > \ldots \lambda_m > \lambda_{m+1} >
 \ldots
 \lambda_E = \frac{p_p}{E} >0, &&  (0 < p_r \leq 1)
\\
 \lambda_1 =1- \frac{p_r}{E}, \;\;
 \lambda_2 =1- \frac{2p_r}{E} - \frac{2p_p}{E^2}   \; .
\eea
The first consequence of these solutions is that the system evolves
to a unique equilibrium solution given by \cite{Evans06}\tnote{Note
that $\Etilde=\Ktilde$ is an interesting limit.  Then all values are
zero except $k=E$ where $p(1)=1$. This is the condensation
limit.}\tnote{We could now give results for the degree distribution
in three limits of particular interest: a) $p_r=1$ the Watts and
Strogatz model \cite{WS98}, b) $p_r=0$ the pure preferential
attachment model and c) $E \ra \infty$ large system or thermodynamic
limit. }
\bea
 G(z) &:=& \lim_{t\rightarrow \infty} G(z,t) =c_0 F(a,b;c;z)  \; .
  \label{Gzresult}
\eea
The time scale for the decay of each of the eigenfunctions
is given by
\beq
 \tau_m = - 1/\ln(\lambda_m) \; .
 \label{taudef}
\eeq

\section{The Degree Distribution}\label{secDD}

\begin{figure}
\includegraphics[width=7.0cm]{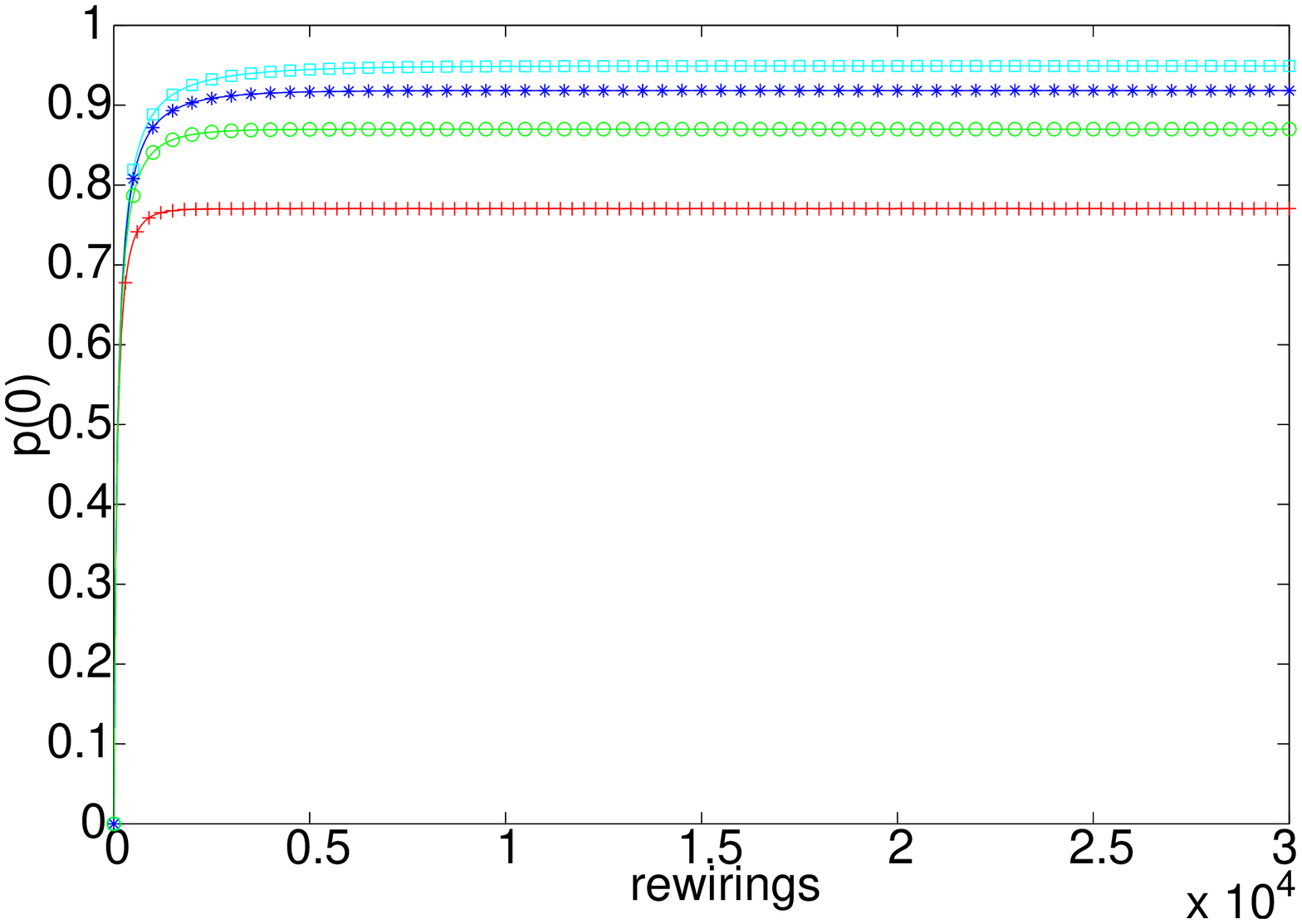}
\includegraphics[width=7.0cm]{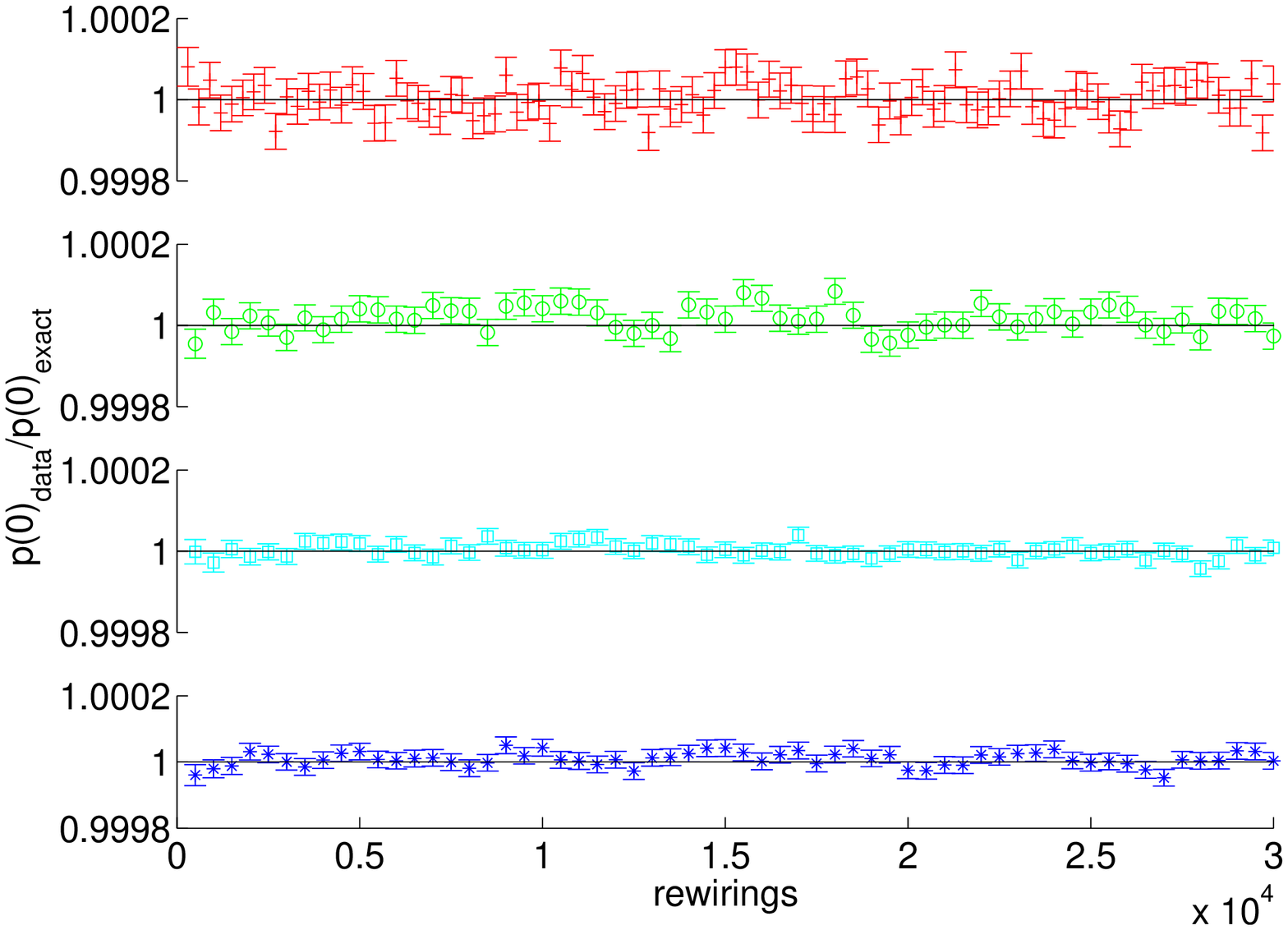}
\caption{Plots of $p(0)$ as a function of
re-wirings and the fractional difference between the simulation
and mean field results for $N=E=100$ and $p_r=0.1$ (crosses),
$p_r=0.04$ (circles), $p_r=0.02$ (stars) and $p_r=0.01$ (squares).
Simulations started with $n(k=1)=E$ and zero
otherwise. Averaged over $10^6$ runs. The solid lines are the results from
the mean field calculations.}
\label{fig:nzero}
\end{figure}

The degree distribution $n(k,t)$ at any time is given as the
coefficients of $z^k$ in the expression for the generating function
$G(z,t)$ of \tref{Gkmdef} and this in turn depends on the initial
conditions. The equilibrium degree distribution derived from $G(z)$
of \tref{Gzresult} which from \tref{Gkmdef} is based only on the
zero-th eigenfunction $\efunc^{(0)}$ (the $m=0$ case of
\tref{Gmresult}). In particular the $k=0$ case shows that $c_0=n(0)
= \lim_{t\rightarrow \infty} n(k=0,t)$ and so we have\tnote{$n(0)$
is the equilibrium value for the number of artifacts with zero edges
attached, $n(k=0,t)=G(z=0,t)$.}
\bea
 n(k)
 &:=& \lim_{t\rightarrow \infty} n(k,t)
  =\frac{1}{k!} \left. \frac{d^kG(z)}{dz^k}\right|_{z=0}
  =\frac{ n(0) }{\Gamma(k+1)}
 \frac{\Gamma(a+k)} {\Gamma(a)} \frac{\Gamma(b+k)}{\Gamma(b)}
  \frac{\Gamma(c)}{\Gamma(c+k)} \; .
\eea
The total number of artifacts is given simply by the generating
function at $z=1$ so
\beq
 N = c_0 G(z=1) = n(0) F(a,b;c;1) \; .
  \label{Nvalueeq}
\eeq
This gives us the equilibrium artifact degree probability
distribution function $p(k)=n(k)/N$ as
\bea
p(k) &=& A \;
 \frac{\Gamma\left( k + \Ktilde \right) }{\Gamma \left(k+1\right) }
  \frac{\Gamma\left( \frac{E}{p_p} - \Ktilde  - k\right) }{\Gamma \left(E  +1  -  k  \right) }
  \, ,
      \label{pkcomplete}
      \\
      A &:=&
      \frac{\Gamma\left(\Etilde\right)\Gamma\left(E+1\right)}{\Gamma\left(\frac{p_r}{p_p}(E-\kav)\right)\Gamma\left(\Ktilde\right)\Gamma\left(\frac{E}{p_p}\right)}
      \, ,
\eea
where we have chosen to write the expression in terms of $\Gamma$
functions of positive arguments and in terms of the original
parameters. Two useful values are the degree probability distribution for zero
degree and maximum degree $k=E$. The former provides a measure of
the number of unused artifacts and thus another measure of the
uniformity of the system and the latter will be discussed in detail
below. These satisfy simple formulae
\bea
 p(0) &=& \frac{\Gamma(\Etilde)}{\Gamma(\Eopp)}
          \frac{\Gamma(\Eopp-\Ktilde)}{\Gamma(\Etilde-\Ktilde)}
           \label{p0value}
\\
 p(E) &=& \frac{\Gamma(\Etilde)}{\Gamma(\Eopp)}
          \frac{\Gamma(E+\Ktilde)}{\Gamma(\Ktilde)} \, .
          \label{pEvalue}
\eea
The results for $p(0,t)$ are plotted against exemplary data in Fig.\
\ref{fig:nzero}.\tnote{$p(E)$ plot too? Plot each in units of its
decay time scale $\tau_2=-\ln(\lambda_2)$.  Make $\ln(p(k))$ the y
axis to show curves clearly.  Try plotting $\ln(f(t+1)-f(t))$ vs.
$t/\tau_2$ for ANY time dependent plot to show up leading time
behaviour.  May not work well.}\tnote{What about calculating the
rank or at least the degree of the largest vertex?}

\subsection{Large Degree Equilibrium Behaviour}

The solution for $p(k)$ has two significant parts. The first $k$
dependent ratio of Gamma functions in \tref{pkcomplete} for $k \gg
1$ and $p_r \approx 0$ behaves as
\beq
 R_1 =
 \frac{\Gamma( k + \frac{p_{r}}{p_p}\kav ) }
      {\Gamma (k+1 ) } \propto k^{-\gamma}
       \left( 1+ O( k^{-1}, \frac{p_{r}\kav}{p_p k}  ) \right),
\qquad
 \gamma = 1- \frac{p_{r}}{p_p}\kav \leq 1 \; .
 \label{gammasimp}
\eeq
For $p_r=0$ or $\kav =0$ (which includes when $N \rightarrow
\infty$) this term gives us an exact inverse $k$ power law for all
degree $k$ from this term.  Another special case corresponds to an
attachment probability of $\Pi_A \propto (k+1)$ which is often found
in the literature, for instance \cite{OTH05,OYT05,OYT06}. This ratio
$R_1$ is then exactly one for all $k$ so $\gamma=0$. In general the
power is usually close but always less than one.

However the $(1-\Pi_A)$ and $(1-\Pi_R)$ terms in \tref{neqngen} have
led to the second $k$-dependent ratio of Gamma functions  in
\tref{pkcomplete}.  If $E \gg k$ this gives an exponential cutoff
\bea
 R_2 =
 \frac{\Gamma\left( \frac{E}{p_p} - \frac{p_{r}}{p_p}\kav  - k\right)} {\Gamma (E+1  -  k  ) }
     &\propto & \exp \{ -\zeta k \} ( 1+ O(\frac{k}{E},\frac{\kav}{E}))
      ,
      \\
      \zeta &=& -\ln \left( p_p \right)
      \label{zetasol}
            \\
      &\approx & p_r
      \; \mbox{ if }
      p_r \ll 1 , \;  \kav \ll E \; .
\eea
Within these approximations, this may be expressed in an equivalent
manner which is some times seen in the literature (e.g.\
\cite{CK70,BHS04,HBH04})
\bea
 R_2 =
  \frac{\Gamma\left( \frac{E}{p_p} - \frac{p_{r}}{p_p}\kav  -k\right)}
 {\Gamma (E+1  -  k  ) }
  &\propto & \left( 1 - \frac{k}{E} \right)^{E \bar{\zeta}} ( 1+ O(\frac{k}{E}))
  \\
  \bar{\zeta} &=& \frac{p_r}{p_p}\left(1-\frac{1}{N}\right) - \frac{1}{E} \approx p_r \; .
\eea
While not strictly valid at $k \approx E$ this second form indicates
that there is a change of behaviour for large degree if
$\bar{\zeta}<0$.  In such a case the numerator of this second
$k$-dependent ratio of Gamma functions becomes very large for $k=E$
and we see directly that this happens if $p_r \ll \psharp$, where
$R_2=1$ ($\bar{\zeta}=0$) at $\psharp$:
\beq
 \psharp = \frac{1}{E+1-\kav} \approx \frac{1}{E}(1+ O(\kav E^{-1} ) )\; .
\eeq
At $p_r=\psharp$  we are closest to a pure power law with a power
$\gamma=\gamma_\sharp = 1-(2N)^{-1}$. For the special case where $N
\rightarrow \infty$ so $\kav \rightarrow 0$ we get a perfect inverse
power law at $p_r=(1+E)^{-1}$.\tnote{In fact this second ratio of
Gamma functions becomes equal to one for all $k$ at $p_r=\psharp
\approx E^{-1}$.  At that value of $p_r=\psharp$ we have no cut off
and this is the closest the degree distribution comes to an exact
inverse power law for all degree values.  The power in this case is
$\gamma_\sharp = 1- (2N)^{-1} \approx 1$.  Compare this with the
$\gamma>2$ found for the tail of the degree distribution of growing
networks.}

As $p_r$ drops below this critical value, a spike emerges at $k=E$
from this second $k$-dependent ratio $R_2$ which comes to dominate
the degree distribution at $p_r \rightarrow 0$. The point where the
distribution has become flat at the upper boundary, so $n(E) =
n(E-1)$  defines an alternative critical random attachment
probability $\pstar$ at
\bea
 \pstar &=& \frac{E-1}{E^2+E(1-\kav)-1-\kav} \; ,
 \\
 E \pstar & \approx & 1+\frac{(\kav-2)}{E} \; .
\eea
Either way when $p_r \lesssim 1/E$ the degree distribution will show
a spike at $k=E$.

Overall we see two distinct types of distribution. For large
random attachment rates, $Ep_r \gtrsim 1$, we get a simple inverse
power with an exponential cutoff\tnote{*** More info on gamma
distributions. A gamma distribution is a general type of
statistical distribution that is related to the beta distribution
and arises naturally in processes for which the waiting times
between Poisson distributed events are relevant.}
\beq
 n(k) \propto (k)^{-\gamma} \exp\{-\zeta k\} \; ,
 \qquad
 p_r \gtrsim \frac{1}{E} \, .
 \label{ngammadist}
\eeq
This behaviour is often noted in the literature
\cite{PLY05,KC64,HBH04,BHS04,XZW05,OTH05,EvansMR00,EH05} and the
formulae given there for the power $\gamma$ and cutoff $\zeta$ or
$\bar{\zeta}$ are consistent with the exact formulae given here
given the various approximations used elsewhere. Note that in any
one practical example it will be impossible to distinguish the
power $\gamma$ derived from the data from a value of one.  This is
because to have a reasonable section of power law behaviour we
require $1 \ll \zeta$ but this implies that $p_r$ is small and so
$(\gamma - 1) \ll \kav$. The power drifts away from one as we
raise the random attachment rate $p_r$ towards one but only at the
expense of the exponential regime starting at a lower and lower
degree. Only when the power is very close to one can we get enough
of a power law to be significant.

However as $p_r$ is lowered towards zero we get a change of
behaviour in the exponential tail around $p_r E \approx 1$.  First
we find the exponential cutoff $\zeta^{-1}$ moves to larger and
larger values, eventually becoming bigger than $E$. For $p_r$
slightly below $\psharp$, that is for $p_r
> \pstar \approx E^{-1}$, the tail starts to rise. For $p_r
E \ll 1$, i.e.\ if there has been no random artifact chosen after
most edges have been rewired once, then we will almost certainly
find one artifact linked to most of the individuals, $n(E) \approx
1$.

This behaviour for $Ep_r \ll 1$ has been noted in some of the
literature where it is known as \emph{condensation}
\cite{BBBJ00,BBJ97,BBJ99,BDJKNPZ02,EvansMR00,EH05} or, in the older
population genetics literature, it is called \emph{fixation}.  It
mirrors similar behaviour known for growing networks when non-linear
attachment probabilities or vertex fitness are used, for example see
\cite{DM01,KRL00}.

\subsection{Limiting Values of $p_r$ in Equilibrium}

The $p_r=0$ limit offers some simplifications as well as being the
only value in the condensation phase.\tnote{Can we do the time
dependence and eigenfunctions in this case?} The condensation is
clear from the expression for $p(k)$ of \tref{pkcomplete} as the
denominator is infinite if $(p_r/p_p)E(1-N^{-1})=0$, i.e.\ either we
have the trivial example of one artifact $N=1$ or we are in the pure
preferential attachment limit $p_p = 1, p_r=0$. For the latter
$p(k)$ is therefore zero for all values of $k$ except $k=0$ or $k=E$
where the infinity cancelled by the same term in the numerator. For
instance we find that\footnote{In this $p_r=0$ limit we have
degeneracy between eigenfunctions number zero and one. However we
will see below that eigenfunction one  $\efunc^{(1)}$ does not
contribute to any physical solution so there is no ambiguity about
our solution in this case.} for small $p_r$ the equilibrium
distribution has the form
\bea
p(0) &\approx & \left(1-\frac{1}{N}\right)
 \left( 1 - p_r \kav [\psi(E) -\psi(1)] \right) + O((p_r)^2)
\\
 p(k) &\approx& p_r \frac{\kav (E-\kav)}{k(E-k)} +O(p_r^2), \qquad
0<k<E
\\
p(E) &\approx & \frac{1}{N}
 \left( 1 - p_r (E-\kav) [\psi(E) -\psi(1) ]\right)
 + O((p_r)^2)
\eea
so only at $p_r=0$ do we get condensation for any $E$. This
represents a true phase transition in the large system ($E \ra
\infty$, thermodynamic) limit between the gamma distribution
\tref{ngammadist} for $p_r>0$ and the condensation $p(k) =
\delta_{k,E}$ at $p_r=0$.\tnote{Can we calculate the
eigenfunctions in these limits?  Both $p_r=0$ and perhaps $E \ra
\infty$?}

At the other extreme, we have the limit of pure random artifact
selection $p_r=1$.  In this limit the model captures exactly the
degree distribution of the original Watts and Strogatz model
\cite{WS98}. In this case for any $E$ and $\kav$ the solution for
$p(k)$ \tref{pkcomplete} reduces to a binomial distribution with a
probability $(1/N)$ of any one edge connecting to a given artifact
vertex, i.e.\ we have the expected Erd\H{o}s-R\'eyni random graph in
the long time limit.\tnote{Can we calculate the eigenfunctions in
this limit?}

\subsection{Time dependence of the Degree Distribution}\label{ssectdep}

So far we have looked at the equilibrium behaviour but we have a
complete solution for the degree distribution for all times and any
value of the parameters through our eigenfunctions \tref{omegamdef}
and eigenvalues \tref{eq:evalues}.  Alternatively for small values
of $E$ it may be more convenient to cast this as a matrix problem
\cite{EP06ECCS}. We have used the latter to predict the degree
distribution for any time for a range of $p_r$ values either side of
and approximately equal to the critical value $\pstar$ in
Fig.s~\ref{fig:halflife0_1}, \ref{fig:halflife0_01} and
\ref{fig:halflife0_001}.  These (and other figures below) show that
the degree distribution evolves on time scales $\tau_2$ set by
eigenvalue number two whatever $p_r$ we use (why it is not $\tau_1$
is explained below). Again the exact mean field results fit the
averaged values from a simulation extremely well.

\begin{figure}[htb]
\includegraphics[width=7.0cm]{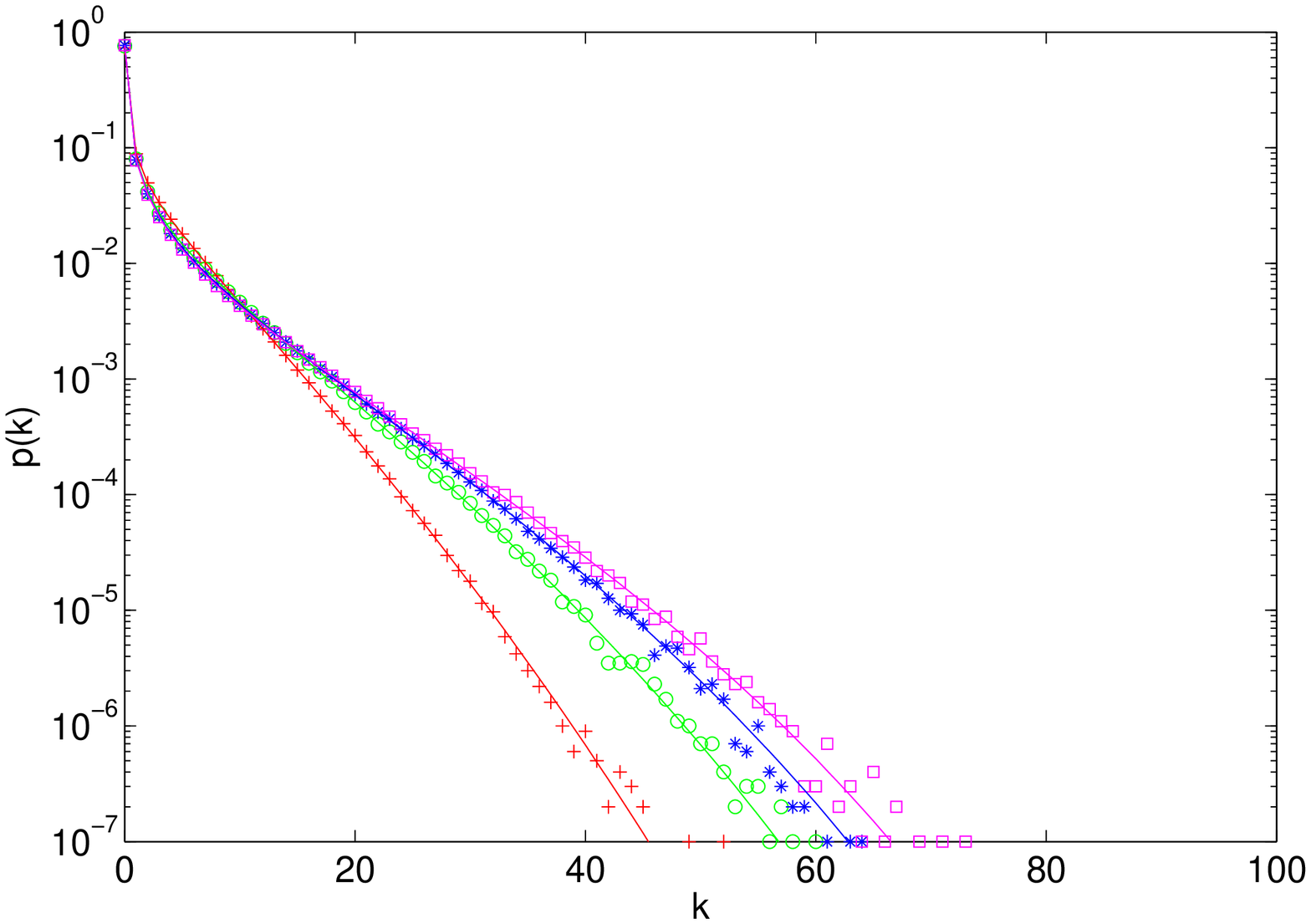}
\includegraphics[width=7.0cm]{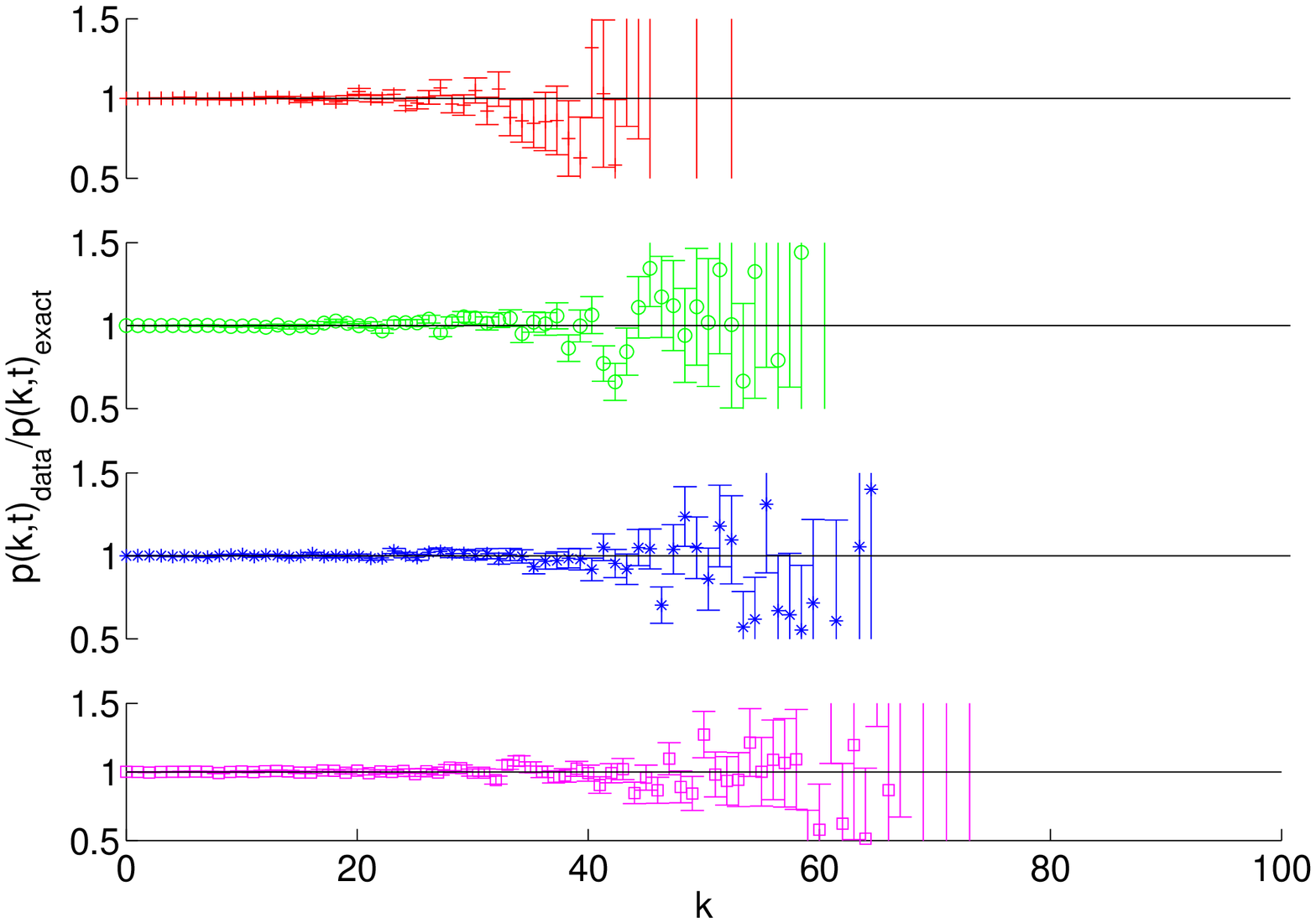}
\caption{Plots of $p(k)$ and the fractional deviation
between the simulation (data points) and exact mean field results
(lines) for $E=N=100$ and $p_r=0.1$ after evolving for
$t\approx\tau_2$ (crosses), $t\approx 2\tau_2$ (circles), $t\approx
3\tau_2$ (stars) and to equilibrium (squares). The solid lines are
the relevant mean field results plotted for the same times. Started
with $n(k=1)=E$ and zero otherwise and simulation results averaged
over $10^5$ runs.  }
 \label{fig:halflife0_1}
\end{figure}

\begin{figure}[htb]
\includegraphics[width=7.0cm]{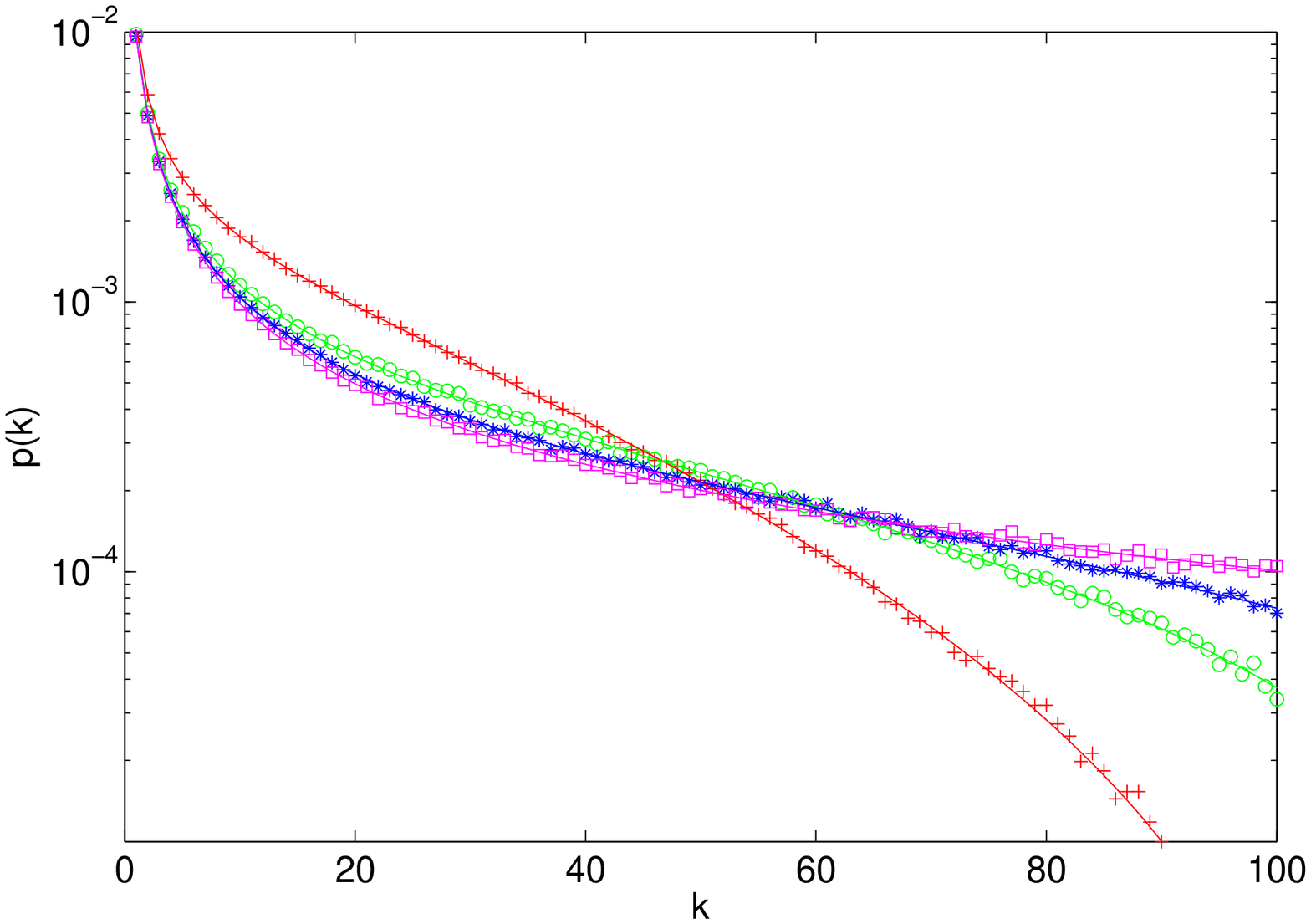}
\includegraphics[width=7.0cm]{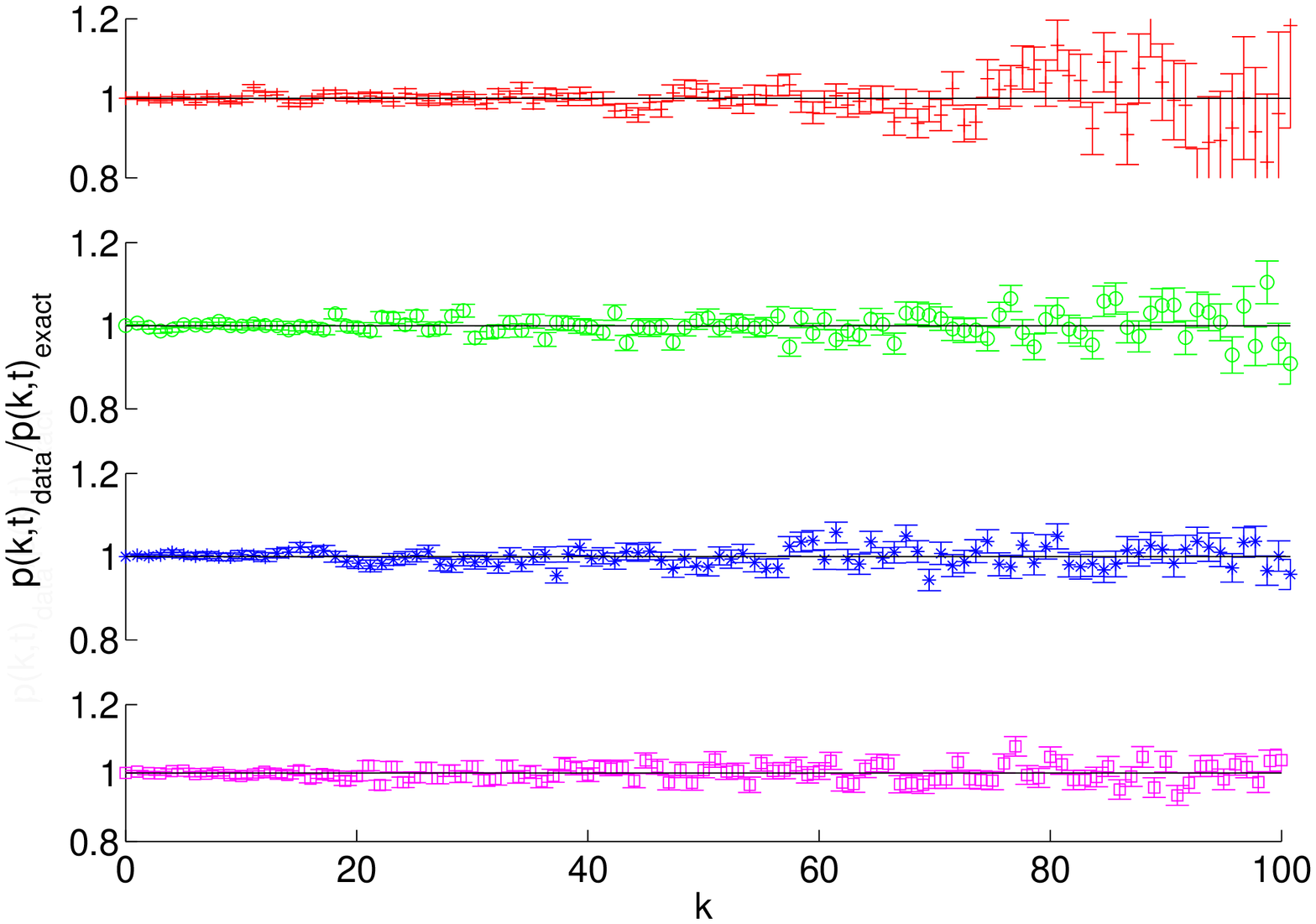}
\caption{Plots of $p(k)$ and the fractional deviation
between the simulation (data points) and exact mean field results
(lines) for $E=N=100$ and $p_r=0.01$ after evolving for
$t\approx\tau_2$ (crosses), $t\approx2\tau_2$ (circles),
$t\approx3\tau_2$ (stars) and to equilibrium (squares). The solid
lines are the relevant mean field results plotted for the same
times. Started with $n(k=1)=E$ and zero otherwise and simulation
results averaged over $10^5$ runs.  } \label{fig:halflife0_01}
\end{figure}

\begin{figure}[htb]
\includegraphics[width=7.0cm]{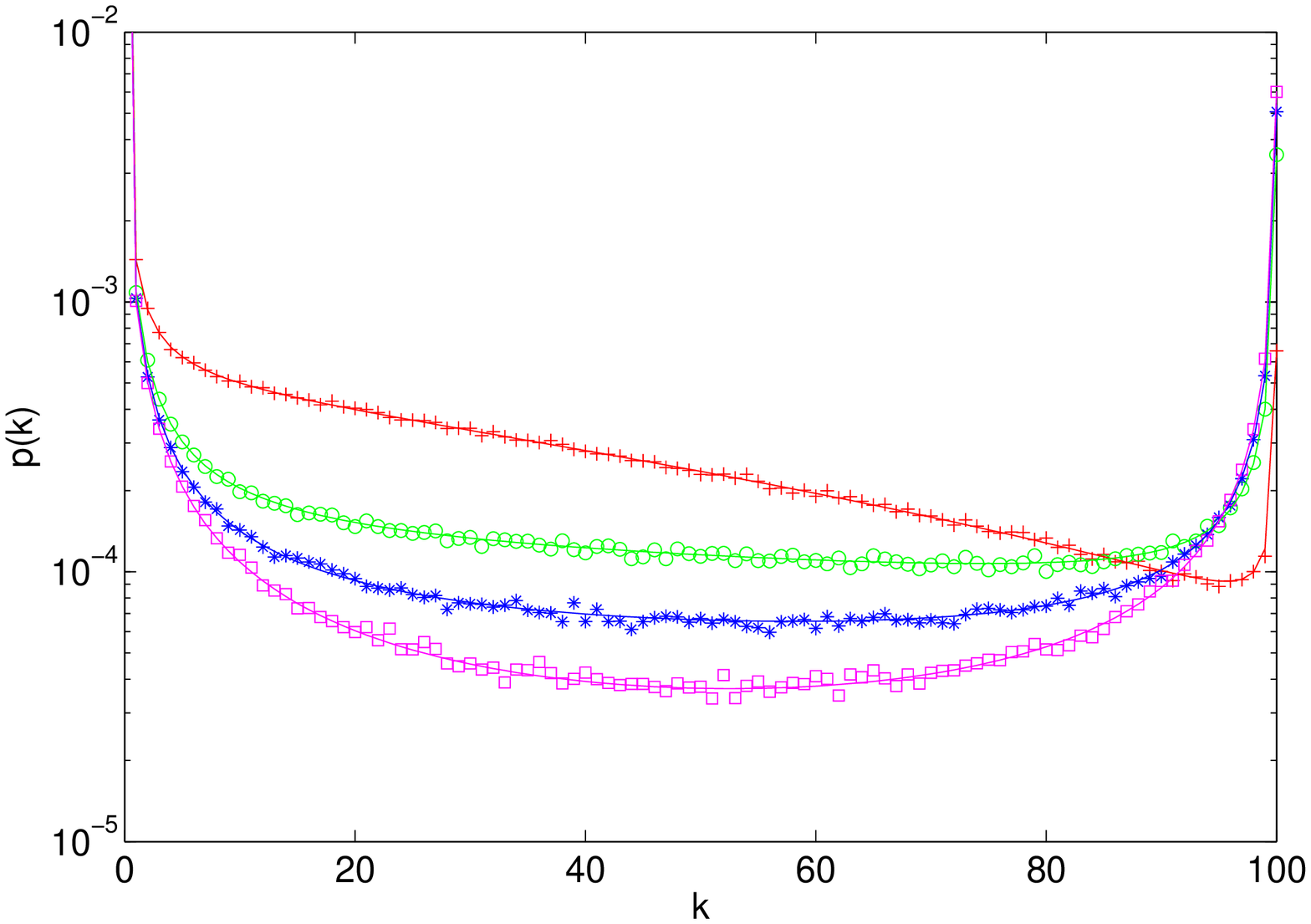}
\includegraphics[width=7.0cm]{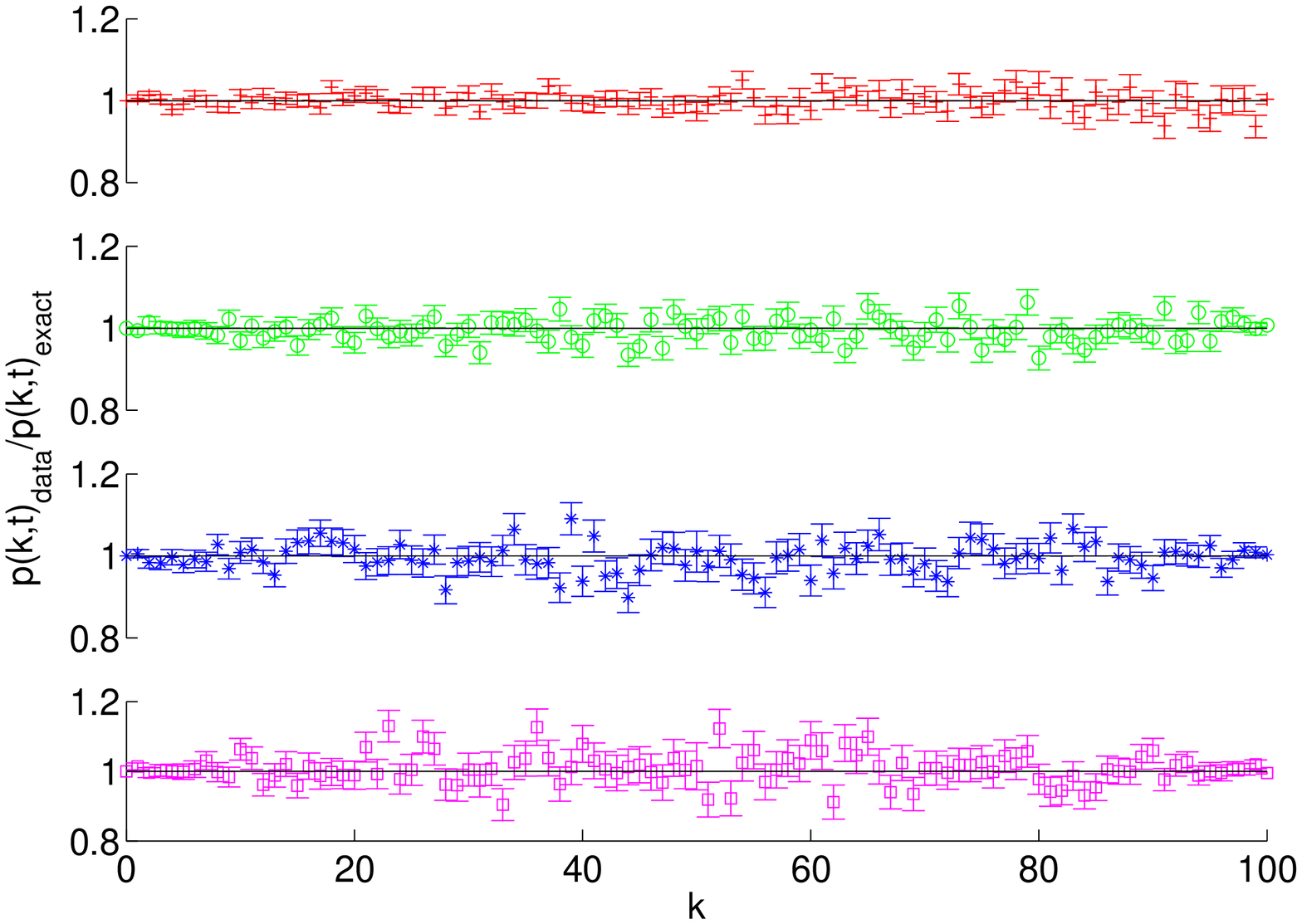}
\caption{Plots of $p(k)$ and the fractional deviation
between the simulation (data points) and exact mean field results
(lines) for $E=N=100$ and $p_r=0.001$ after evolving for
$t\approx\tau_2$ (crosses), $t\approx2\tau_2$ (circles),
$t\approx3\tau_2$ (stars) and to equilibrium (squares). The solid
lines are the relevant mean field results plotted for the same
times. Started with $n(k=1)=E$ and zero otherwise and simulation
results averaged over $10^5$ runs.} \label{fig:halflife0_001}
\end{figure}

\section{The Moments of the Degree Distribution}\label{secmoments}

\begin{figure}
\includegraphics[width=7cm]{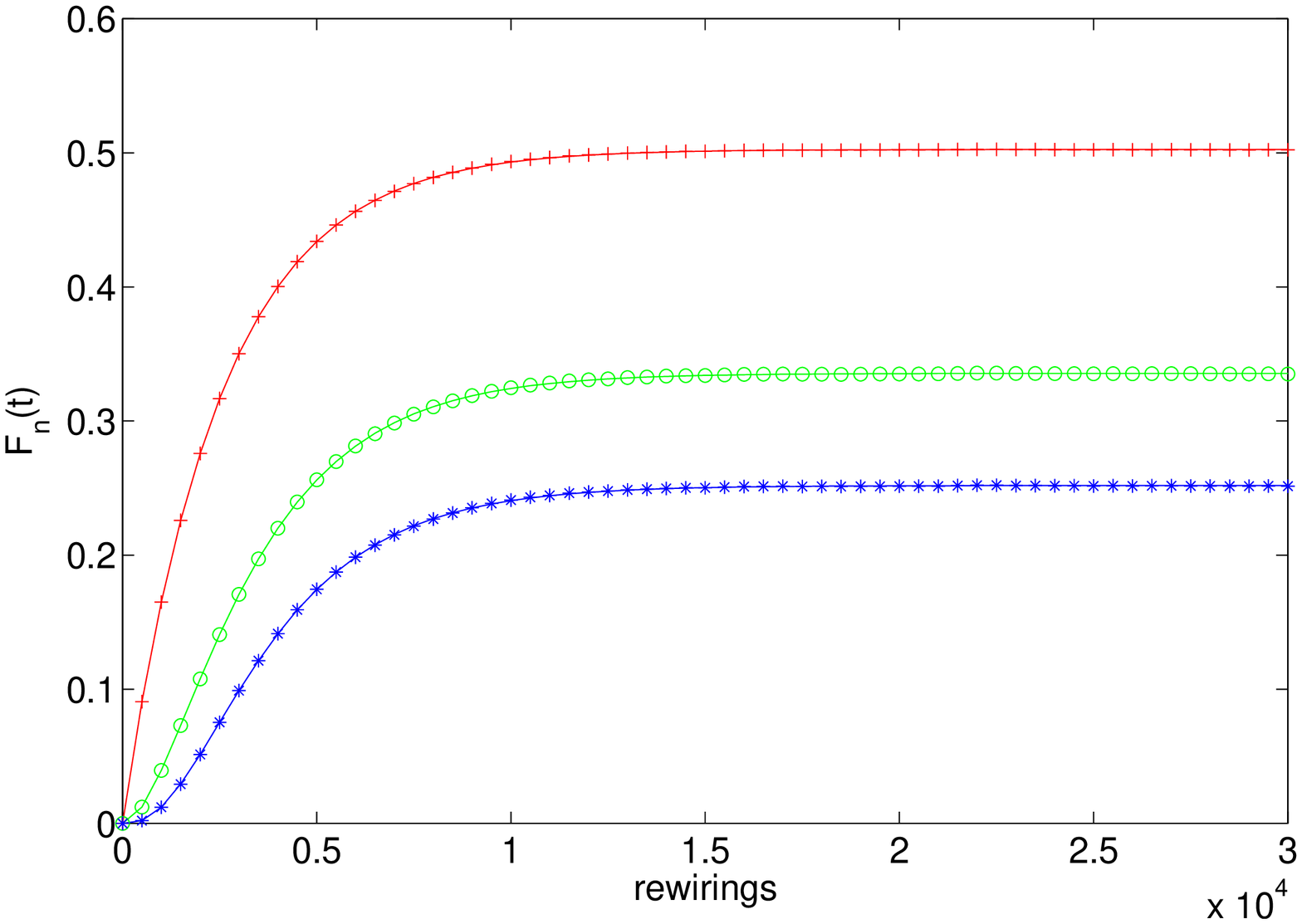}
\includegraphics[width=7cm]{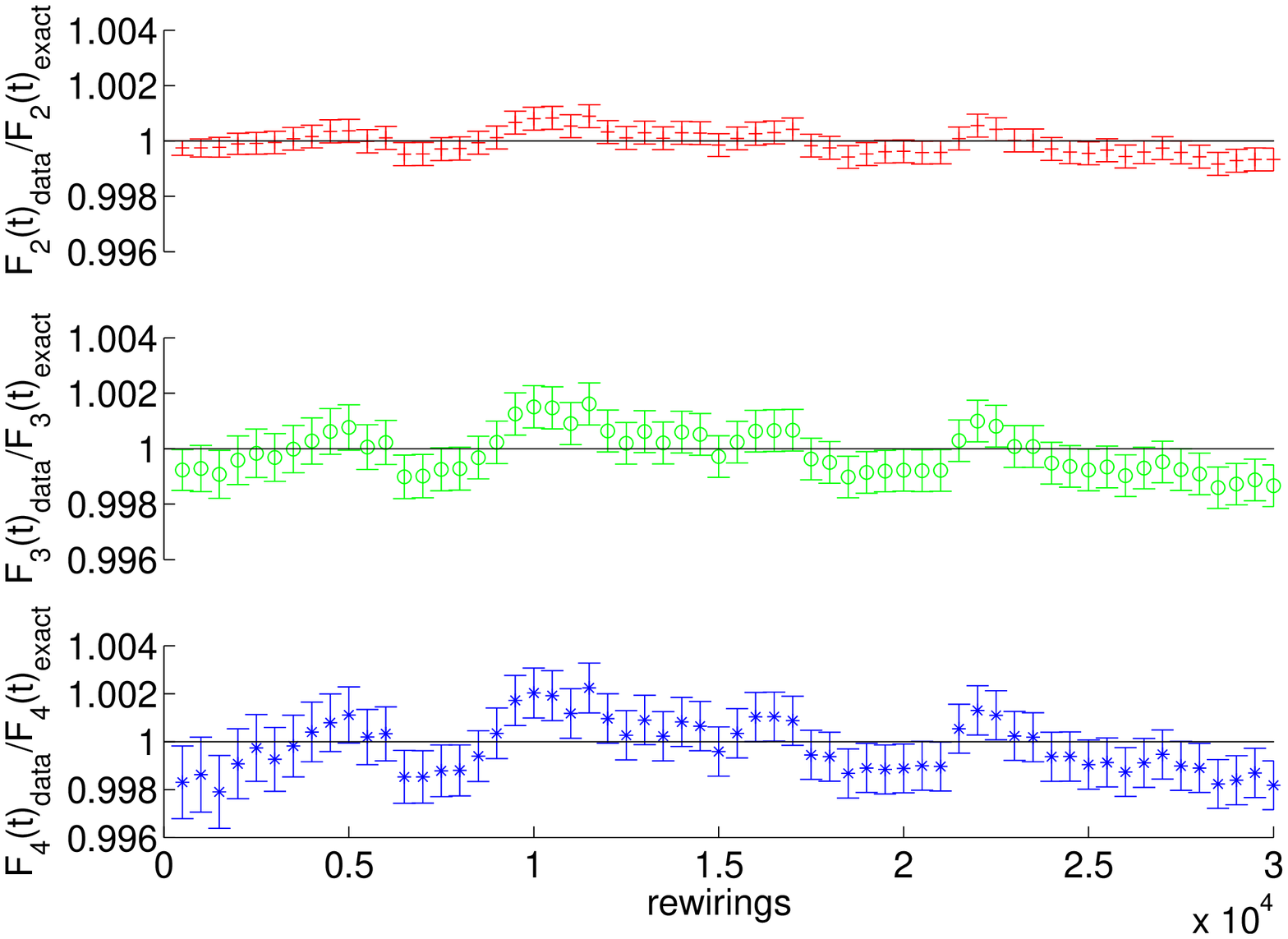}
\caption{Plots of various $F_n(t)$ (points) with their exact mean
field predictions (lines).  From top to bottom we have: $F_2(t)$
(crosses), $F_3(t)$ (circles), $F_4(t)$ (stars). For $E=N=100$,
$p_r=0.01$ and data points are the average of $10^5$ runs of a
simulation.  } \label{fig:fnVarious}
\end{figure}

The properties of the hypergeometric function mean it is easy to
calculate derivatives of the generating function at $z=1$ at
\emph{any} time as we have
\begin{eqnarray}
 g^{(m)}_n &:=& \left.\frac{d^n G^{(m)}(z)}{dz^n}\right|_{z=1}
 \label{gmndef}
 \\
 &=&
 (-1)^m g^{(0)}_0
 \frac{\Gamma(n+1)}{\Gamma(n+1-m)} \frac{\Gamma(a+n)}{\Gamma(a+m)}
 \frac{\Gamma(b+n)}{\Gamma(b+m)}
 \nonumber \\
 && \times \frac{\Gamma(c-n-m-a-b)}{\Gamma(c-a-b)} \frac{\Gamma(c-a)}{\Gamma(c-a-m)}
 \frac{\Gamma(c-b)}{\Gamma(c-b-m)}
 \qquad (m \leq n)
 \label{moments}
\end{eqnarray}
with $g^{(m)}_n =0$ if $m>n$. This then suggests that rather than
work in terms of the higher moments $\langle k^n \rangle$, we use
the probabilities $F_n$ where\footnote{This means that the
generating function may be written as $G(z) = \sum_{n=0}^{E} (z-1)^n
\binom{E}{n}F_n$, a form useful when solving the differential
equation \tref{eq:gendiffm}.}
\beq
 F_n(t) :=   \frac{\Gamma(E+1-n)}{\Gamma(E+1)}
 \left. \frac{d^nG(z,t)}{dz^n}\right|_{z=1} =
  \sum_{k=0}^E \frac{k}{E}\frac{(k-1)}{(E-1)} \ldots
 \frac{(k-n+1)}{(E-n+1)} n(k,t)
 \; .
 \label{Fndef}
\eeq
The function $F_n$ is the probability that if we choose $n$ distinct
edges, they will all share the same artifact. The $r$-th moment
$\taverage{k^r}$ can be calculated if given all the $F_n$ for $n
\leq r$. The $F_n$ achieve their largest value only when we have a
condensation, $p(k)=(N-1)\delta_{k,0}+\delta_{k,E}$ where $F_n=1$
for all $n \geq 2$. That is we have a perfectly homogeneous
population (all the individuals are connected to the same artifact).
The lowest possible value of $F_n$ depends on the other parameters.
If we have $E \leq N$ then when all artifacts have at most one edge
attached then $F_n = 0$ for all $n$, as we can see in
Fig.~\ref{fig:fnVarious} at the initial time.  The evolution causes
a drift towards a more heterogeneous distribution. The same Figure
also shows how the exact mean field results match results from
simulations extremely well. Mathematically it is clear from the
result \tref{Gmresult} that the $F_n$ only has contributions from
the first $n+1$ eigenfunctions, i.e.\ from $G^{(m)}$ for $m \leq
n$.\tnote{Does $G^{(m)}$ have an interpretation in terms of coming
from vertices which have m edges
--- already, at least, at most, or something?}

In equilibrium only eigenfunction zero contributes and we have a
simple result\tnote{*** ALTERED BY REMOVING
$\Gamma(1+E)/\Gamma(1+E-n)$ FACTOR ***}
\bea
  \lim_{t \ra \infty} F_n(t) := F_n
  &=&
   N\frac{\Gamma(\Ktilde+n)\Gamma(\Etilde)}
      {\Gamma(\Ktilde)\Gamma(\Etilde+n)} \; .
      \label{Fneq}
\eea

\subsection{Normalisation $N$}\label{ssecN}

The zero-th moment sets the overall normalisation of the degree
distribution $n(k,t)$.  This is nothing but the total number of
artifact nodes $N$ and for any time $t$ we find it is equal to
\beq
 N = G(z=1,t) = \sum_{m=0}^E c_m (\lambda_m)^t g^{(m)}_0
   = c_0  F(a,b;c;1) \; .
 \label{Nvalue}
\eeq
This result is time independent because it comes only from the
zero-th eigenvector, the only time independent part of the solution.
Thus our solution is consistent with a key property in this model,
namely the constant number of artifacts $N$.  This then fixes the
amplitude of the zero-th eigenfunction in \tref{Gkmdef} to be
\beq
 c_0 = \frac{N}{g^{(0)}_0} \; .
 \label{c0value}
 \eeq

\subsection{Average Degree}\label{sseckav}

The first derivative of the generating function gives the number of
edges
\bea
 E
 &=&  \left. \frac{d}{dz} G(z,t)\right|_{z=1}
 \label{Evalue}
 \\
 &=&  \sum_{m=0}^E c_m (\lambda_m)^t g^{(m)}_1
 = \frac{c_0}{N} g^{(0)}_1 + \frac{c_1}{N} (\lambda_1)^t g^{(1)}_1 \; .
\eea
Only eigenfunctions zero and one contribute but the latter leads to
time-dependence.  On the other hand we also have a fixed number of
edges in this model as it is one of our input parameters. The only
solution is therefore $c_1=0$. Thus for any physical problem there
is no contribution from the eigenfunction number one.

This equation then appears to over constrain our solution as $c_0$
is already known from the normalisation \tref{c0value} and all other
quantities are fixed. However, we find that using standard
properties of hypergeometric functions and the normalisation from
\tref{Nvalue} that the solution already satisfies \tref{Evalue} and
again everything is consistent.

\begin{figure}
\includegraphics[width=7.0cm]{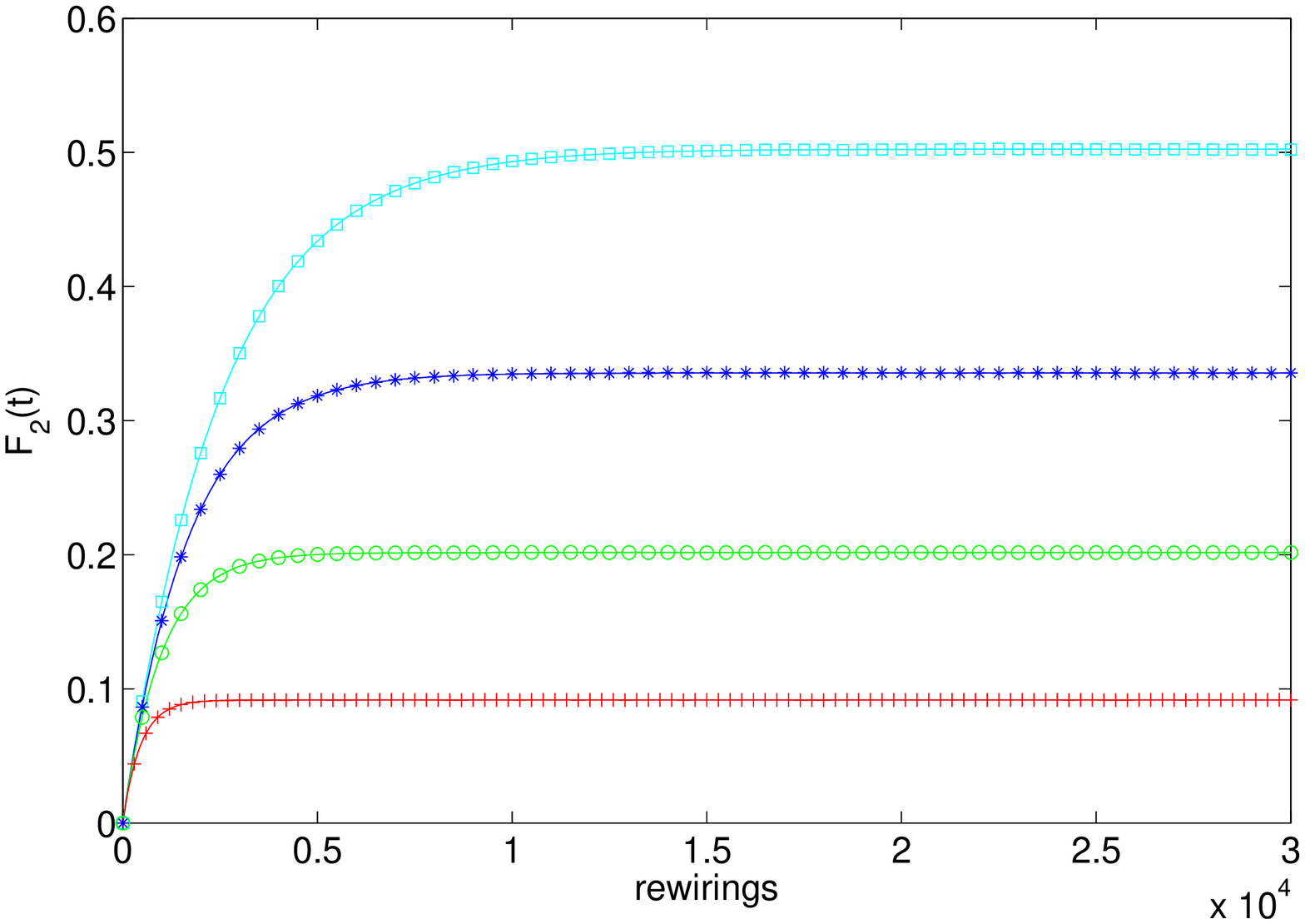}
\includegraphics[width=7.0cm]{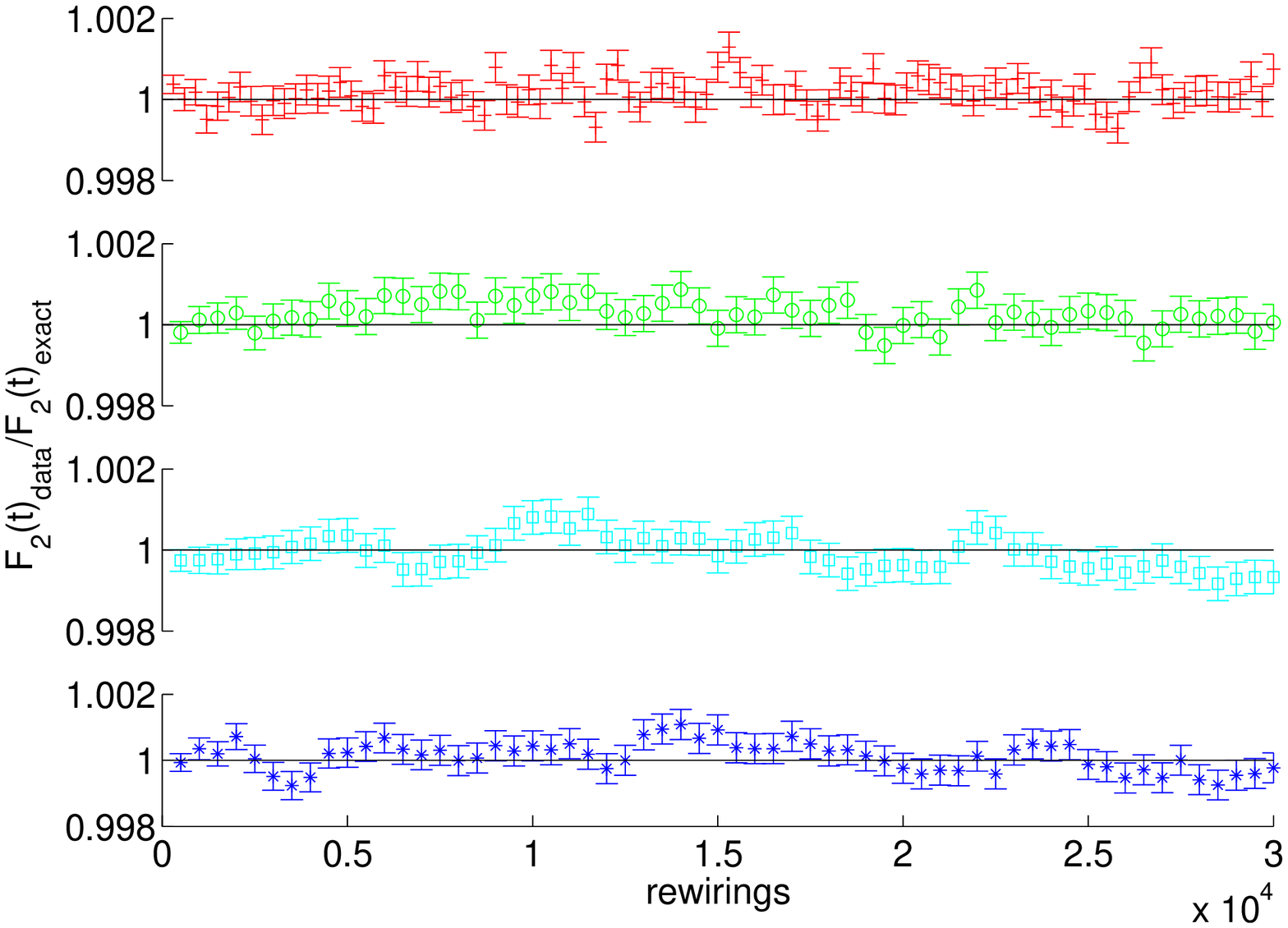}
\caption{Plots of the Homogeneity Factor $F_2(t)$ and the
fractional difference between the simulation (data points) and
exact mean field results (lines) for $N=E=100$ and different
$p_r$. From bottom to top: $p_r=0.1$ (crosses), $p_r=0.04$
(circles), $p_r=0.02$ (stars) and $p_r=0.01$ (squares). Initial
configuration is $n(k=1) = E$ and zero otherwise. Simulation data
is averaged over $10^4$ runs. The results are in good agreement
with the analytic result equation \tref{eq:homogeneity}. }
\label{fig:Homogeneity}
\end{figure}

\subsection{Homogeneity Measures $F_n$ and Initial Conditions}\label{ssecF2}

The next derivative of the generating function contains the second
moment but it is preferable to work with our related function $F_2$
-- the probability that two distinct edges chosen at random are
connected to the same artifact. Similar measures of the homogeneity
of the artifact choices have been used before such as $F =
\taverage{(k^2/E^2)}$ but this is easily calculated from our $F_2$
measure. We find that
\bea
 F_2(t) &:=& \sum_{k=0}^E \frac{k(k-1)}{E(E-1)} n(k,t)
  \label{eq:F2tdef}
 \\
 &=&
 \frac{1}{E(E-1)}
 \left(
  c_0 g^{(0)}_2 + c_2 (\lambda_2)^t g^{(2)}_2
  \right)
  \label{eqF2tres}
\eea
Now there is time dependence but only coming from eigenfunction
number two. This function is readily evaluated using
(\ref{moments}), the coefficient $c_2$ fixed upon specification of
the initial conditions.  This formula fits the results extremely
well as Fig.~\ref{fig:Homogeneity} shows.\tnote{Extract $\lambda_2$
from the data?}

One of the advantages of the $F_n$ measures is that they provide a
systematic and practical way of fixing the amplitudes of each
eigenfunction, the $c_m$ coefficients of \tref{Gkmdef}, from the
initial conditions. From the definition \tref{Fndef} we can express
$F_n$ in terms of the $n$-th derivatives of the generating functions
associated with the $m$-th eigenfunction evaluated at $z=1$, i.e.\
$g^{(m)}_n$ of \tref{gmndef}.
\begin{equation}
\frac{\Gamma(E+1-n)}{\Gamma(E+1)} \sum_{m=0}^{n}c_m
g_n^{(m)}=F_n(t=0) .
 \label{Fncnreln}
\end{equation}
However only the first $n$ eigenfunctions contribute so we can use
an iterative scheme to find the first few coefficients quickly.
These are sufficient to provide an excellent approximation for the
degree distribution for most times.

For example consider the case of uniform initial conditions, such
that $E\leq N$ and each artifact is connected to at most one edge,
then we have that $F_n=0$ for $n \geq 2$.   This corresponds to the
choice of initial conditions used in obtaining the numerical results
in Figures~\ref{fig:nzero} though \ref{fig:halflife0_001}. So for
these initial conditions the following condition holds,
\begin{equation}
\sum_{m=0}^{n}c_m g_n^{(m)}=0,\qquad\qquad n \geq 2.
\end{equation}
We have already seen that the $N$ parameter fixes $c_0$ in
\tref{c0value} while the first moment or equivalently $E$ gives
$c_1=0$. So starting with $n=2$ we have
\begin{equation}
c_2=-c_0\frac{g_2^{(0)}}{g_2^{(2)}}.
\end{equation}
The exact time dependence of the second Homogeneity function is
\begin{eqnarray}\label{eq:homogeneity}
F_2(t)&=&(1-\lambda_2^t) F_2(\infty) \nonumber \\
&=&(1-\lambda_2^t)\frac{p_p+p_r\langle k\rangle}{p_p+p_r E}
\end{eqnarray}
Comparisons to numerical results are plotted in
Fig.~\ref{fig:Homogeneity}.

Another particularly convenient choice of initial conditions is to
attach each individual vertex to the same artifact vertex so that
$n(k=E)=1$, $n(k=0)=N-1$ and zero otherwise. Now $F_n(0)=1$ and the
condition \tref{Fncnreln} becomes
\begin{equation}
\sum_{m=0}^{n}c_m g_n^{(m)}
 =
 \frac{\Gamma(E+1)}{\Gamma(E+1-n)} .
\end{equation}
Note, we have put no restriction on the total number of individual
vertices, $E$. For the simplest case $n=2$ we are led to another
simple formula,
\begin{equation}
F_2(t)
 =
 \left(1-\lambda_2^t\right)
 \left( \frac{p_p+p_r\langle k\rangle}{p_p+p_r E} - 1 \right)
 + 1.
\end{equation}

\begin{figure}
\includegraphics[width=5.2cm]{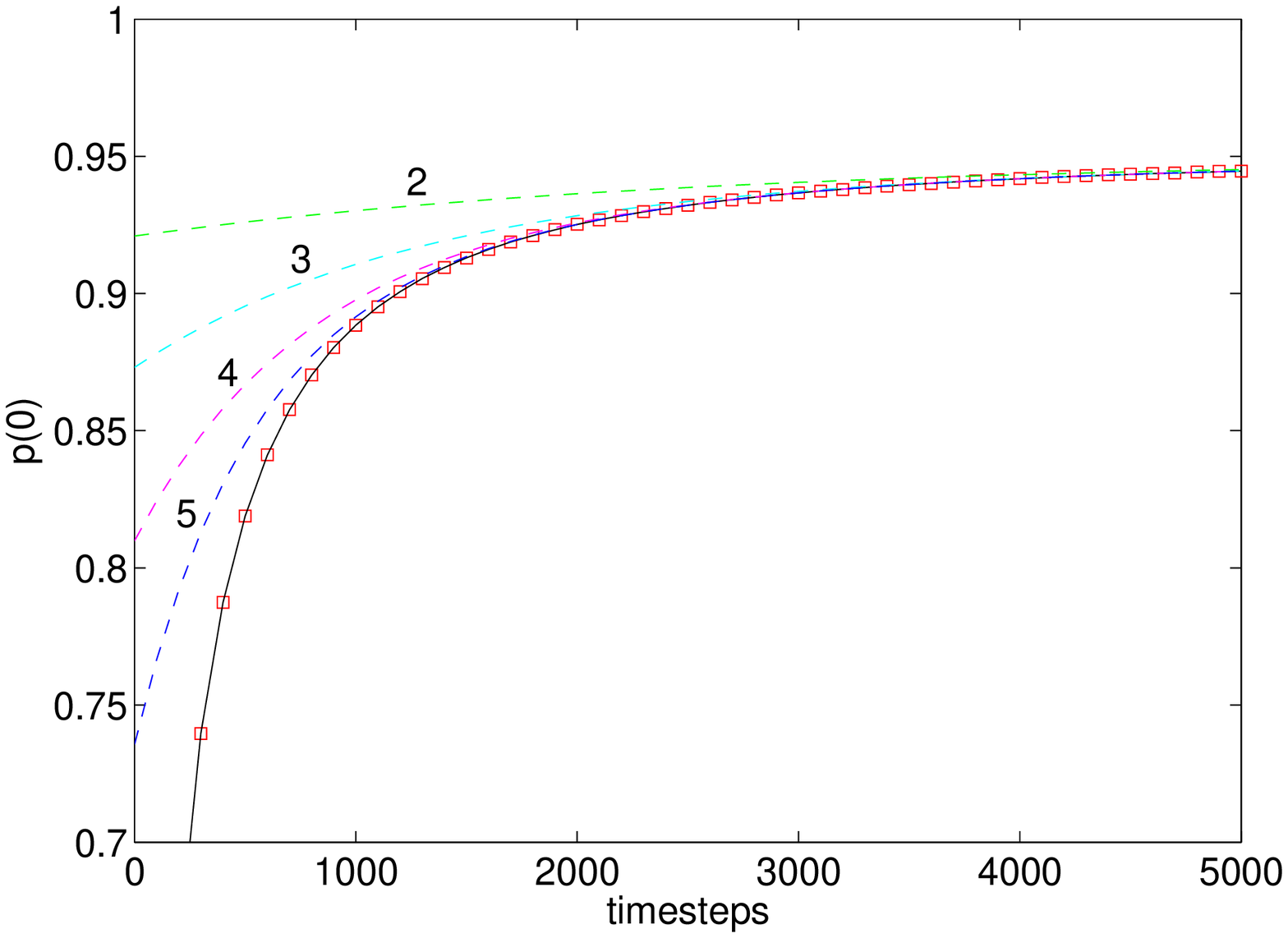}
\includegraphics[width=5.2cm]{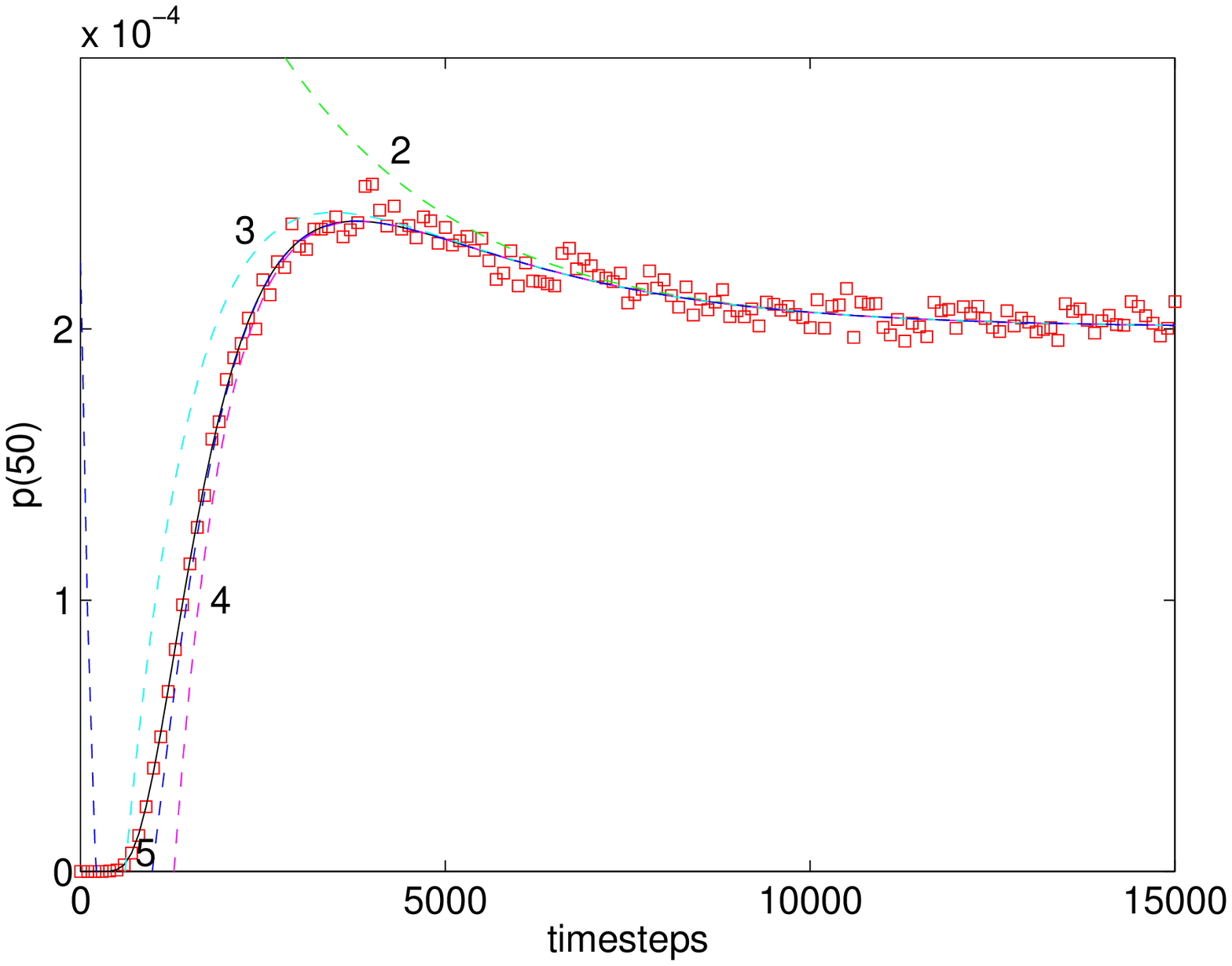}
\includegraphics[width=5.2cm]{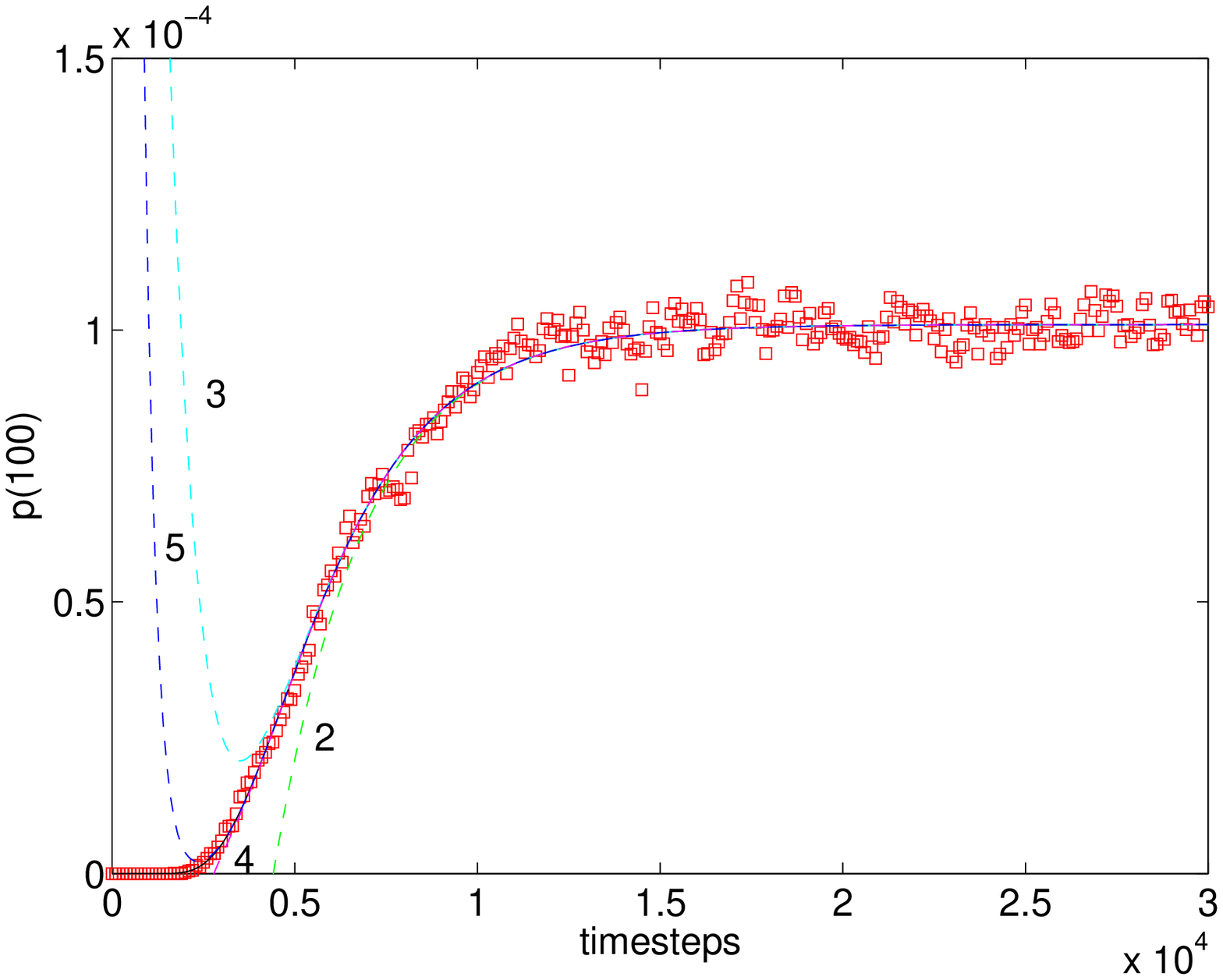}
\caption{Plots of the evolution of $p(0)$, $p(50)$ and $p(100)$ for
$E=N=100$, $p_r=0.01$ compared with the relevant mean field results.
The solid lines represent the exact mean field solution, while the
numbered dashed lines indicate the successive improvements obtained
using contributions from $\lambda_2$ (2) up to $\lambda_5$ (5).
Simulations started with $n(k=1)=E$ and zero otherwise. Averaged
over $10^5$ runs. } \label{fig:Contrib}
\end{figure}

\begin{figure}
\includegraphics[width=5.2cm]{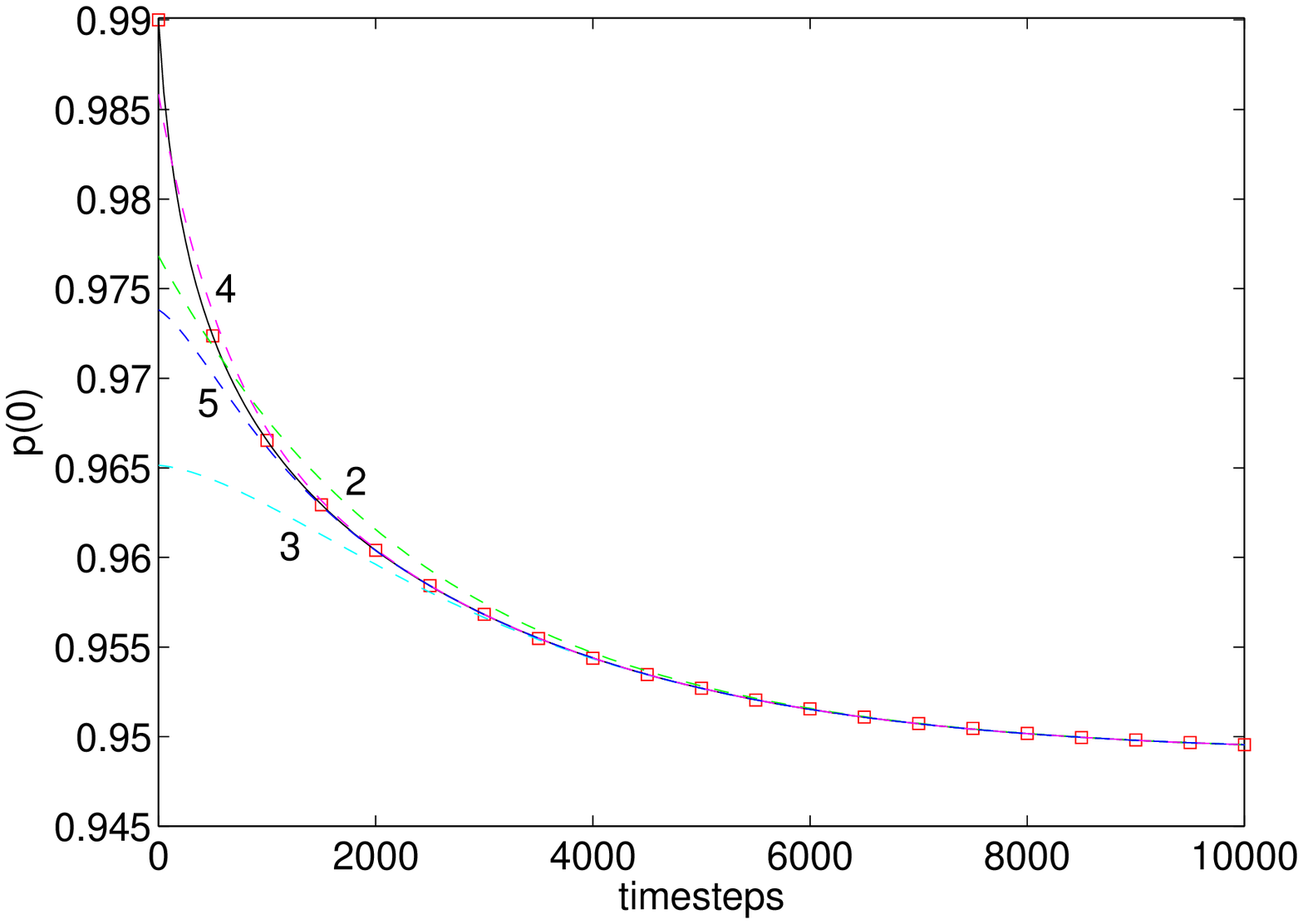}
\includegraphics[width=5.2cm]{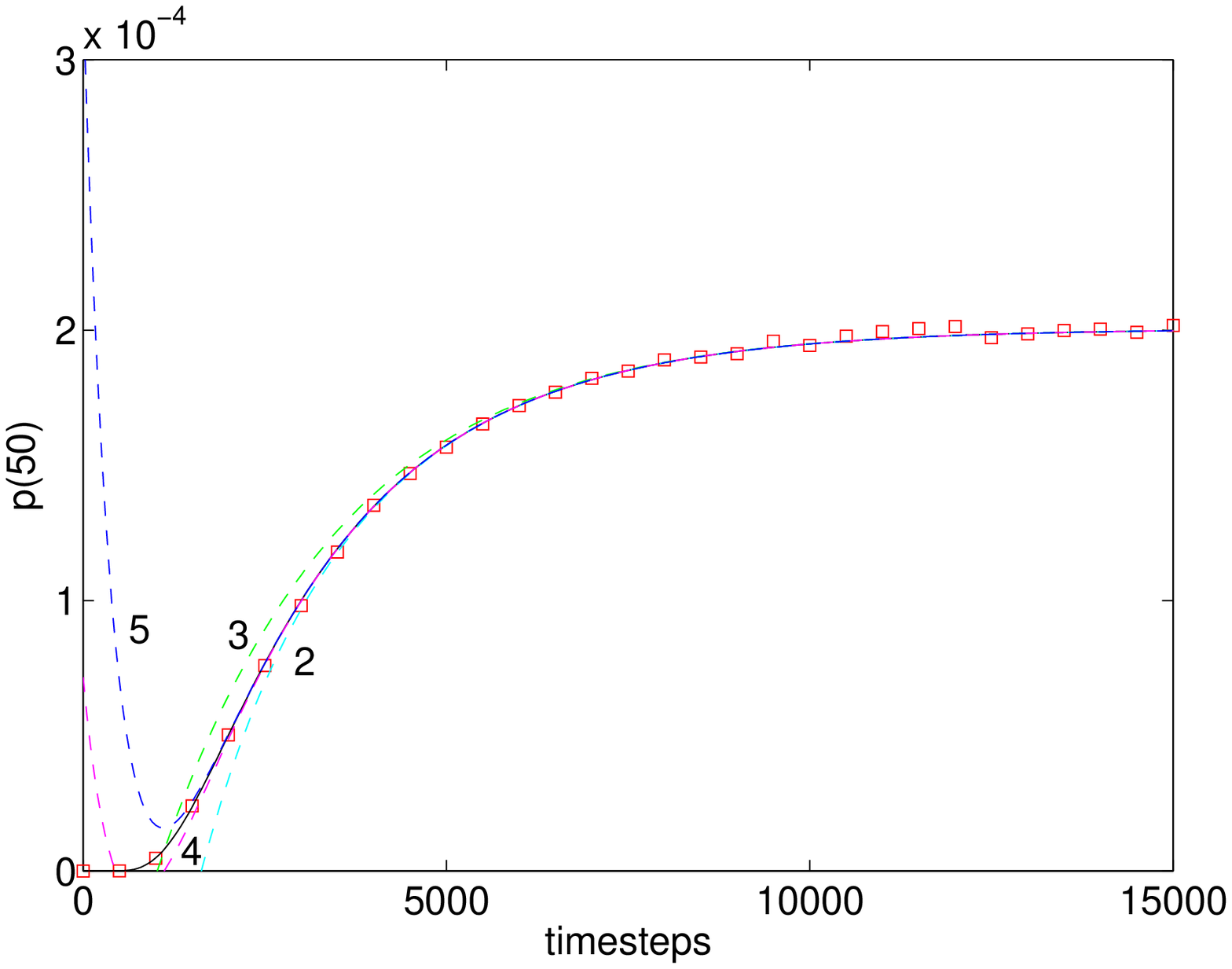}
\includegraphics[width=5.2cm]{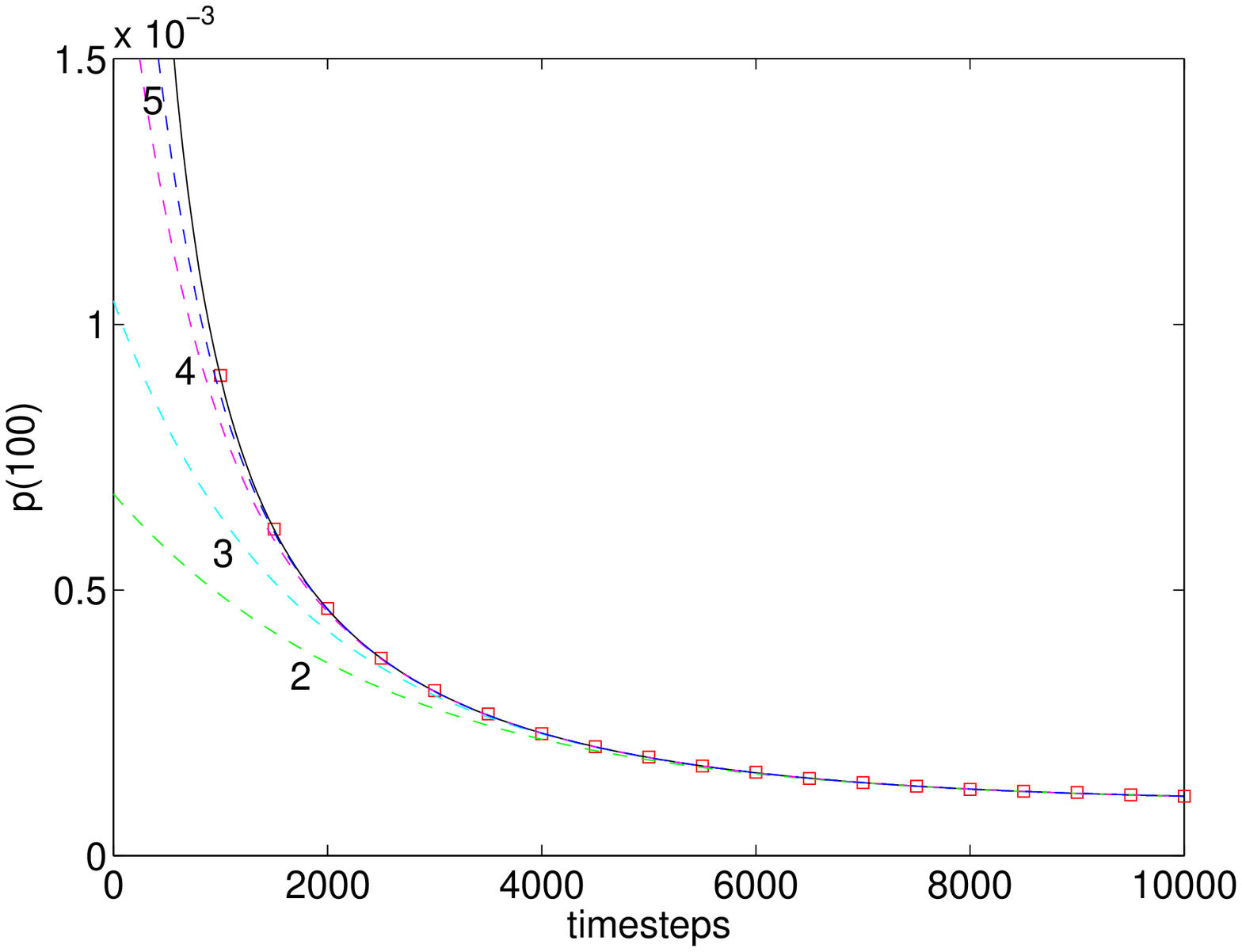}
\caption{Plots of the evolution of $p(0)$, $p(50)$
and $p(100)$ for $E=N=100$, $p_r=0.01$ compared with the relevant
mean field results. The solid lines represent the exact mean field
solution, while the numbered dashed lines indicate the successive
improvements obtained using contributions from $\lambda_2$ (2) up to
$\lambda_5$ (5).
Simulations started with the alternate initial condition $n(k=E)=1$.
Averaged over $10^6$ runs. } \label{fig:ContribAltIC}
\end{figure}

While the exact degree distribution requires knowledge of all the
eigenfunctions, the low $n$ eigenfunctions still provide a suitable
approximation for most time. Fig.s \ref{fig:Contrib} and
\ref{fig:ContribAltIC} illustrate this, showing the contributions to
the degree distribution from successive eigenvectors for the two
initial conditions discussed above.

\tpre{\newpage}
\section{Discussion of other Models}\label{discussion}

The bipartite network of Fig.\ref{fCopyModel2} represents
relationships at the core of many models in the literature, some of
which are not usually expressed in terms of networks.  While the
models considered elsewhere often have additional elements compared
with our simple model of Sec.\ \ref{sec:model}, those models often
contain special cases where the degree distributions for any time
will be given by our exact result.  The aim of this section is to
indicate the relationship between our model and those found
elsewhere.  Only some of these connections have been made before and
then only in some of the literature.  We will start by considering
some generalisations of our simple model as this will then help us
to make comparisons with previous work.

\subsection{Generalisations of our bipartite model}\label{ssgenmodel}

Many related models work in continuous time so that the number of
rewiring events which have occurred corresponds to our discrete time
variable.  However it is easy to recast our simple model as a
continuous time process so that $n(k,t+1)-n(k,t)$ on the left hand
side of our master equation \tref{neqngen} becomes
$\mathrm{d}n(k,t)/\mathrm{d}t$ with the $\Pi_R$ and $\Pi_A$ being
interpreted as rates. This case is just as easy to solve as we
replace the form used before for our degree distribution and
generating function \tref{Gktdef} by
\beq
 n(k,t) = \sum_{m=0}^{E} c_m   \omega^{(m)}(k) \exp\{-\bar{\lambda}_m t \}
 \; ,
 \qquad
 \bar{\lambda}_m = 1-\lambda_m \; .
 \label{conttime}
 \eeq
The eigenfunctions $\omega^{(m)}(k)$ and the associated generating
functions $G^{(m)}$ are exactly as before, the new eigenvalues
$\bar{\lambda}_m$ have a simple relationship to our original
$\lambda_m$ and the form of the time dependence is altered. Thus our
exact solutions may be applied to discrete or continuous time.

Another obvious generalisation of our model is to alter the form of
the attachment and removal probabilities $\Pi_R$ and $\Pi_A$
\tref{PiRPiAsimple}. Suppose
\bea
 \Pi_R(k) &=& q_p\frac{k}{E} + q_a\frac{(1-\delta_{k,0})}{N_a(t)}
         \\
 \Pi_A(k) &=&
             p_p\frac{k}{E} + p_a\frac{(1-\delta_{k,0})}{N_a(t)} + p_r\frac{1}{N}
 \, ,
 \label{PiRPiAfull}
\eea
with $q_p + q_{a} =1$ and $p_p + p_{a} +p_r=1$.  We have added an
extra process to our model where with probability $p_a$ ($q_a$) we
can attach (remove) an edge from a random artifact chosen uniformly
from those which have at least one edge.  The number of such
\emph{active} artifacts, those where $k>0$, is denoted $N_a(t)$ and
this is time-dependent.  However the master equation \tref{neqngen}
will no longer be exact because $N_a$ varies from configuration to
configuration so averages of ratios of $k^m n(k)$ and $N_a$ will not
factorise into the ratio of their averages \tref{nonfactor}. The
time dependence of $N_a(t)$ also makes the non-equilibrium solutions
of the master equation hard to find though the equilibrium solution
can be found as before \cite{Evans06}.  For instance the slope of
the power law section in the non-condensed phase is now
\beq
\gamma = 1- \left( \frac{p_{r}}{p_p} + \frac{p_{a}}{p_p} -
\frac{q_{a}}{q_p}\right) \kav
 \label{gammasimpgen}
\eeq
and it can now be greater than one. In terms of the condensed phase,
this now occurs when $p_p > q_p$ which means that there is now a
range of parameter values which lead to this phase for large
networks.

Another obvious generalisation is to have terms in $\Pi_A$ or
$\Pi_R$ proportional to general powers of the degree
$(k^\beta/z_\beta)$ or powers of general functions $(a + b
k)^\beta$.\tnote{Could talk about Doug's results for $0<\beta<1$ and
$\beta>1$ and relate to some of the literature for asymptotic
links.} As noted in Section \ref{sec:model} this means that the
master equation \tref{neqngen} is then only an approximation though
in many cases it will be a good one.\tnote{and it is not clear how
such attachment probabilities could arise in a real system.}

An important aspect of our model is that we have events where an
edge is rewired back to the same artifact so that the configuration
does not change.  We had to include the $(1-\Pi)$ factors to account
for this correctly.  If we wish we can exclude these events which
corresponds to choosing an attachment probability of the form
\beq
 \Pi_A(d,a) =
 \left[ \tilde{p}_r\frac{1}{(N-1)}
     +  \tilde{p}_p\frac{k_a}{E-k_d}\right] (1-\delta_{d,a}),
   \qquad \tilde{p}_p+\tilde{p}_r=1
 \label{PiAnotad}
\eeq
where we are removing an edge from artifact labelled $d$ (the
departure artifact) and adding it to an artifact labelled $a$ (the
arrival artifact). The $(1-\Pi)$ factors in the master equation
\tref{neqngen} are now always one and can be dropped, giving the
master equation the form often seen in the literature (e.g.\ in
\cite{GBM95,GL02,DM03,OTH05,OYT06,PLY05,XZW05}). However the
preferential attachment term $\tilde{p}_p$ of the attachment
probability $\Pi_A$ now has a configuration dependent normalisation.
The mean-field master equation is now an approximation for all
$\tilde{p}_p>0$ and it is hard to solve it for arbitrary times. In
many cases the fluctuations will be small and the mean field will be
a good approximation. Further if the number of edges attached to any
one artifact is small (tends to zero in the large $E$ limit) then
the difference between our model and one excluding $a=d$ events will
be small \cite{DM03}. Unfortunately, this will not be true in the
interesting case where we have a condensation since for some
artifact vertices $k_a/E$ will be significant, finite even in the
large $E$ limit. We would then expect differences between processes
based on \tref{PiRPiAsimple} and \tref{PiAnotad}.

Ultimately we could make the attachment or removal rates depend on
the individual nature of each vertex, e.g.\ make the probabilities
$p_r$ and $p_p$ vary with artifact vertices. This could mimic
`fitness' where some artifacts are intrinsically more likely to
attract edges.

One realistic way that artifact fitness could emerge is through the
addition of an Artifact graph. That is we could add a second network
which connects artifacts to artifacts and this could be used in
choosing how the bipartite graph is rewired.  For instance, suppose
we have chosen the edge we are going to rewire, so that we know the
departure artifact. We could choose the arrival artifact by making a
random walk on the artifact graph starting from the departure
artifact \cite{ES05,SK04}. In this way the artifacts with high
degree in the artifact graph would be preferred (even for a walk of
one step) and a natural fitness assignment for artifacts has
emerged. Alternatively, we could view this Artifact graph as a way
of encoding some distance metric on the artifact space.  That is
when choosing a random artifact, a $p_r$ event, it may be that a
small variation in the artifact, as defined by some metric, is more
likely than a large one.  Our simple model is equivalent to having a
complete graph with tadpoles\footnote{A tadpole in a general graph
is an edge starting and ending at the same vertex.  For example see
vertex A in  Fig.\ref{fCopyModel2und}.} for the Artifact graph (the
adjacency matrix is one for all entries) which we use for the random
choice ($p_r$) events. The variation mentioned above, where
reconnection to the same artifact is excluded, corresponds to a
complete Artifact graph with no tadpoles.

At the moment, our preferential attachment process, $p_p$, has been
put in by hand.  However this can emerge naturally if we add an
Individual graph, one which just connects the individual vertices.
Suppose we have chosen the individual whose edge is to be rewired.
We now make a random walk on the Individual graph and arrive at an
individual vertex which is connected to what is now taken to be the
arrival artifact for the rewiring process.  Even a short walk of
this type produces good approximations to preferential attachment
processes \cite{ES05,Vaz00,Vaz02,SK04}.  The preferential attachment
events in our simple model are equivalent to doing a random walk on
an Individual graph which is a complete graph with tadpoles.

\subsection{Relationship to models in the literature}

\begin{figure}[htb]
{\centerline{\includegraphics[width=10cm]{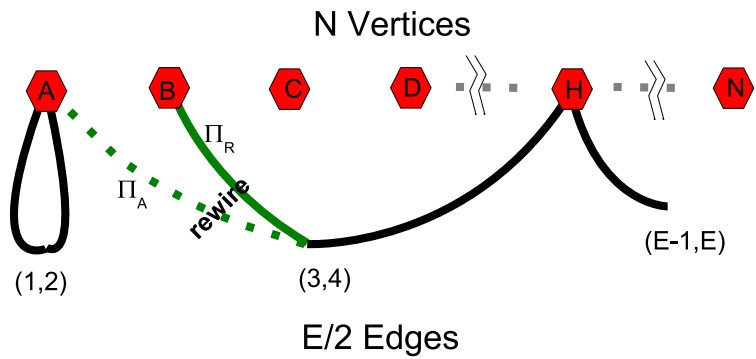}}}
\caption{How the rewiring of the bipartite graph represents the
rewiring of an undirected unipartite graph.  In this example
individual vertices numbers $(2i)$ and $(2i-1)$ are paired in the
bipartite graph to give the edge labelled $(2i-1,2i)$ in the
equivalent undirected graph. The rewiring of the undirected graph
depicted in this figure is is equivalent to that shown for the
bipartite graph rewiring of Fig. \ref{fCopyModel2}. This is the
projection used by Molloy and Reed \cite{MR95}.}
 \label{fCopyModel2und}
\end{figure}

The rewiring of unipartite networks has been studied in its own
right \cite{WS98,BCK01,DMS03,DM03,PLY05,XZW05,OTH05,OYT05,OYT06} but
all of these examples contain, in some sense, our bipartite graph. A
projection of our bipartite graph gives an unipartite graph, made
from just the artifact vertices. One way to achieve this example is
to pair the individual vertices (say individuals numbered $(2i-1)$
with $(2i)$) and to consider the two edges of these individual
vertices as the two ends (the stubs) of a single edge in the new
undirected graph\tnote{This is an example of how a separate
unipartite graph made up of the individual vertices can used to
record more information that the bipartite graph alone can record.}.
Thus the process our simple model illustrated in Fig.\
\ref{fCopyModel2} represents a rewiring process in some undirected
graph as shown in Fig.\ \ref{fCopyModel2und}. In this way, or by
considering the problem directly, we see that the mean field
equations \tref{neqngen} are the same and we need only alter the
normalisations in the probabilities
\tref{PiRPiAsimple}.\tnote{Rewiring of such graphs was considered in
\cite{WS98,XZW05,OYT06}. Fitness \cite{OYT05,OYT06}??? FULL LIST???}

Note that the degree distribution of the projected undirected graph
at any one time is independent of how we pair off individual
vertices in the bipartite graph. Thus the degree distributions of
many different unipartite networks is represented by the same
bipartite graph. Indeed this is the same projection used by Molloy
and Reed to construct general random graphs, that is graphs of a
given degree distribution but otherwise arbitrary
\cite{MR95}.\tnote{This also means that our quantity $F_2(t)$ has
the interpretation that it is ??? in the projected unipartite
graph.}

There are several expressions for global properties of large
generalised random graphs which depend on the ratio of the second
and first moments through a parameter $z$
\cite{MR95,MR98,NSW01,DMS03a}
\beq
 z(t):= \frac{\langle k^2 \rangle}{\langle k \rangle} -1 = (E-1) F_2(t)
\eeq
Thus for large general random graphs being rewired using any mixture
of random vertex and preferential attachment, we can give these
global properties at \emph{any} time.  For instance the mean inter
vertex distance $\ell$ scales as $(\ln(N)/\ln(z(t) +
\mathrm{const})$ so we see from \tref{eq:homogeneity} that for large
$E$ we only avoid $\ell$ scaling as $\ln(N)$ if $p_r E \sim O(1)$.

Similarly a GCC (giant connected component) is present in this
unipartite projection when $z>1$.  We see that this will always
appear if $\kav >1$ or, if $\kav<1$, it appears only if
$p_r<(2-\kav)^{-1}$. Suppose we start from the most disconnected
example where $F_2(t=0)=0$ (so $\kav \leq 1$). Using
\tref{eq:homogeneity} we can find the time at which the GCC first
appears. If $p_r E \sim O(1)$, which includes the condensate region,
we find that the GCC appears at $t=E/2$.  This is much quicker than
the approach to the equilibrium configuration which happens on a
time scale $\tau_2 \sim O(E^2)$. If $p_r$ is raised from $O(E^{-1})$
towards the critical value for the existence of a GCC,
$(2-\kav)^{-1}$, the time at which the GCC appears increases,
reaching infinity at the critical value of $p_r$.

A different example of this projection is when our initial bipartite
graph has each artifact connected to $m$ individuals
($n(k)=N\delta_{km}$).   The unipartite graph projection is then a
randomised version of the graphs used by Watts and Strogatz
\cite{WS98}. From this initial condition and setting $p_r=1$ we
therefore have the exact solution for the degree distribution at any
time in the Watts and Strogatz model. The pairwise correlation of
individual vertices is only required if we want to know about other
aspects of the Watts and Strogatz networks, such as the network
distance and clustering coefficients which were the focus of
\cite{WS98}.  We can though calculate such quantities at any time in
the randomised graph which provides a useful comparison.

It is straightforward to adapt this projection so that we get a
directed graph. For instance the direction of an edge in the
projected unipartite graph could flow from the artifact connected to
individual $(2i-1)$ to the artifact connected to individual $(2i)$.
A simple modification of the master equation \tref{neqngen} is
needed to keep track of the in- and out-degree if we choose to make
these edges directional.

\begin{figure}[htb]
{\centerline{\includegraphics[width=10cm]{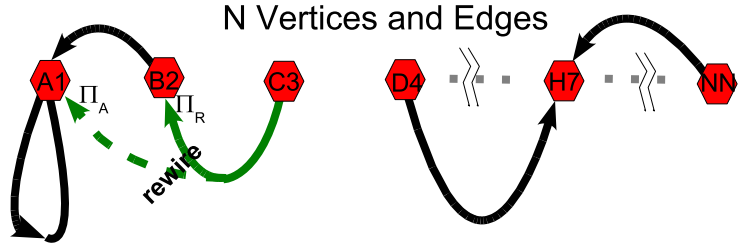}}}
\caption{Another type of projection from our bipartite to a
unipartite network.  Each individual vertex is fused with one of the
$N$ artifact vertices to produce a unipartite graph with $N$
vertices. The edges of the unipartite graph are naturally
directional coming from the vertex associated with the old
individual vertex of the bipartite graph and going to the old
artifact vertex. In the simplest case we have the same number of
artifacts $E$ as individuals $N$ and then we fuse artifact $A$ with
individual $1$ to give a unipartite vertex labelled $A1$ etc. This
produces a network of the type used in \cite{PLY05}.  The figure
shown here is this simple projection of the bipartite graph and
rewiring event of Fig. \ref{fCopyModel2}.}
 \label{fCopyModelPLY}
\end{figure}

One can also think of other types of projection onto unipartite
graphs. Suppose one fuses each individual vertex $i$ with an
artifact vertex which is not necessarily the artifact connected to
that individual by the individual's edge in the bipartite
graph.\tnote{In general artifacts may be fused more than one
individual or with none at all.} The individual-artifact edges of
the bipartite graph now represent edges between the fused vertices
of this projected unipartite graph. There is a natural direction
associated to these unipartite edges coming from the
individual-artifact direction of the bipartite graph, and this can
be maintained or ignored as needed. A simple example of this
projection is where the numbers of individual and artifact vertices
are the same and we fuse each artifact vertex with one individual
vertex. If we let the edges of the bipartite graph represent edges
from the individual to an artifact in the unipartite graph, then the
unipartite graph vertices have out-degree equal to one and this is
one way of representing the degree distribution of the graphs of
\cite{PLY05} and as illustrated in Fig.\ \ref{fCopyModelPLY}. Our
master equation \tref{neqngen} is then the exact mean field
description for the in-degree in this case.

We will now turn to problems where there is no explicit reference to
a network in the standard exposition but which can still be related
to our model. In such cases there is an implicit graph in the
problem which one may define to make contact with our realisation,
but this network may not be relevant in these other studies.

The work on cultural transmission
\cite{Neiman95,BS03,HB03,HBH04,BHS04,BS05} is usually developed
without reference to any network. The names for our vertices come
from this case.  In this context individuals are deemed to be
choosing some artifact of no particular value (pottery designs,
pedigree dog breeds or baby names for example) by \emph{copying} the
choice of another individual --- preferential attachment. Sometimes
though one can expect \emph{innovations} to be made when a
completely new artifact is introduced.  This translates to a random
attachment event in the $N \ra \infty$ limit.  While this work does
not generally use a network picture, it does translate directly into
our network model (e.g.\ see \cite{BS03}).  Here the edges in our
network realisation represents the artifacts chosen by each
individual.  In cultural transmission problems, samples of these
distributions are often available, from records of births, pedigree
dog registrations, or reports from archaeological
excavations\footnote{Sometimes these samples are medium term time
averages of the degree distribution $\int p(k,t)$ and such
distributions may take a different form, a problem addressed in
\cite{HB03,HBH04}.}.

It is relatively easy to see how the same model may be used for
family names rather than the personal names of \cite{HB03}. In this
case the partners who change their family name are represented by
the individual vertices, the family names are the artifact vertices
and the edges represent the partners who keep their family name.
This is then the constant population limit of the models in
\cite{ZM01}.

This family name example shows that this model may be linked to
inheritance processes.  As noted elsewhere
\cite{BS03,HB03,HBH04,BHS04} the oldest examples come from a simple
model for the diversity of genes in a constant population due to
Kimura and Crow \cite{KC64,CK70}. In the case of a haploid cell
(viruses, bacteria and blue-green algae provide examples) the
artifacts are alleles of a single gene carried by each individual.
The preferential attachment events correspond to \emph{inheritance}
of genes.  This produces a \emph{drift} towards homogeneity and, if
unchecked, a condensation or \emph{fixation} in the frequency of
alleles in the population. The random attachment process is
\emph{mutation} in this context.

\begin{figure}[thb]
{\centerline{\includegraphics[width=10cm]{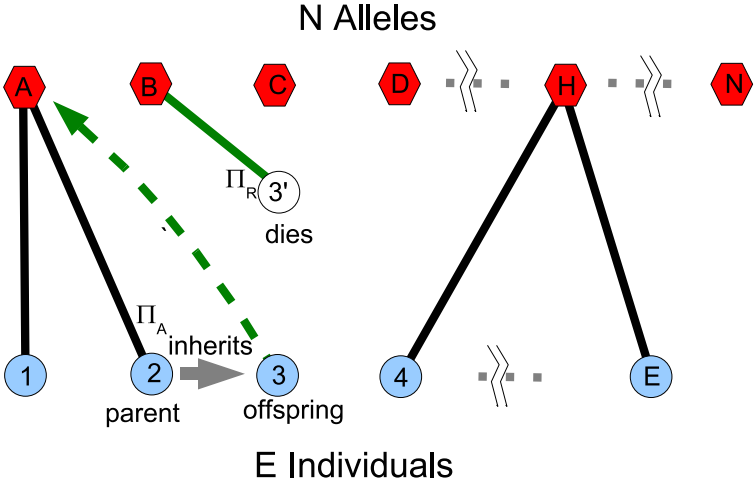}}}
\caption{Interpretation of the the example shown in Fig.
\ref{fCopyModel2} as a haploid gene inheritance and mutation model.
Each individual carries one copy of a gene and each different
version of the gene, an allele, is represented by an artifact
vertex.  The edges indicate the allele present in each individual.
Note this also serves as a model of family names for a constant
population if one partner inherits the family name of the other
partner. In this case the alleles (artifacts) are the family names,
the edges are the males and the genes (individuals) are the
females.}
 \label{fhaploid}
\end{figure}

In general a fitness is assigned to each gene representing the
chance of survival and the successful birth rate associated with
having that gene. There may also be different mutation rates
associated with each gene. However in the simplest models such
factors are ignored. Translating the haploid gene model for a
constant population into the language of our network rewiring model
is then simple.  The organisms are the individuals and we consider
one gene carried by each individual. Each different allele of this
gene is a distinct artifact vertex and so each edge records the
allele carried by an individual. The rewiring example of
Fig.~\ref{fCopyModel2} is translated into a haploid model as shown
in Fig.~\ref{fhaploid}.

\begin{figure}[bht]
{\centerline{\includegraphics[width=10cm]{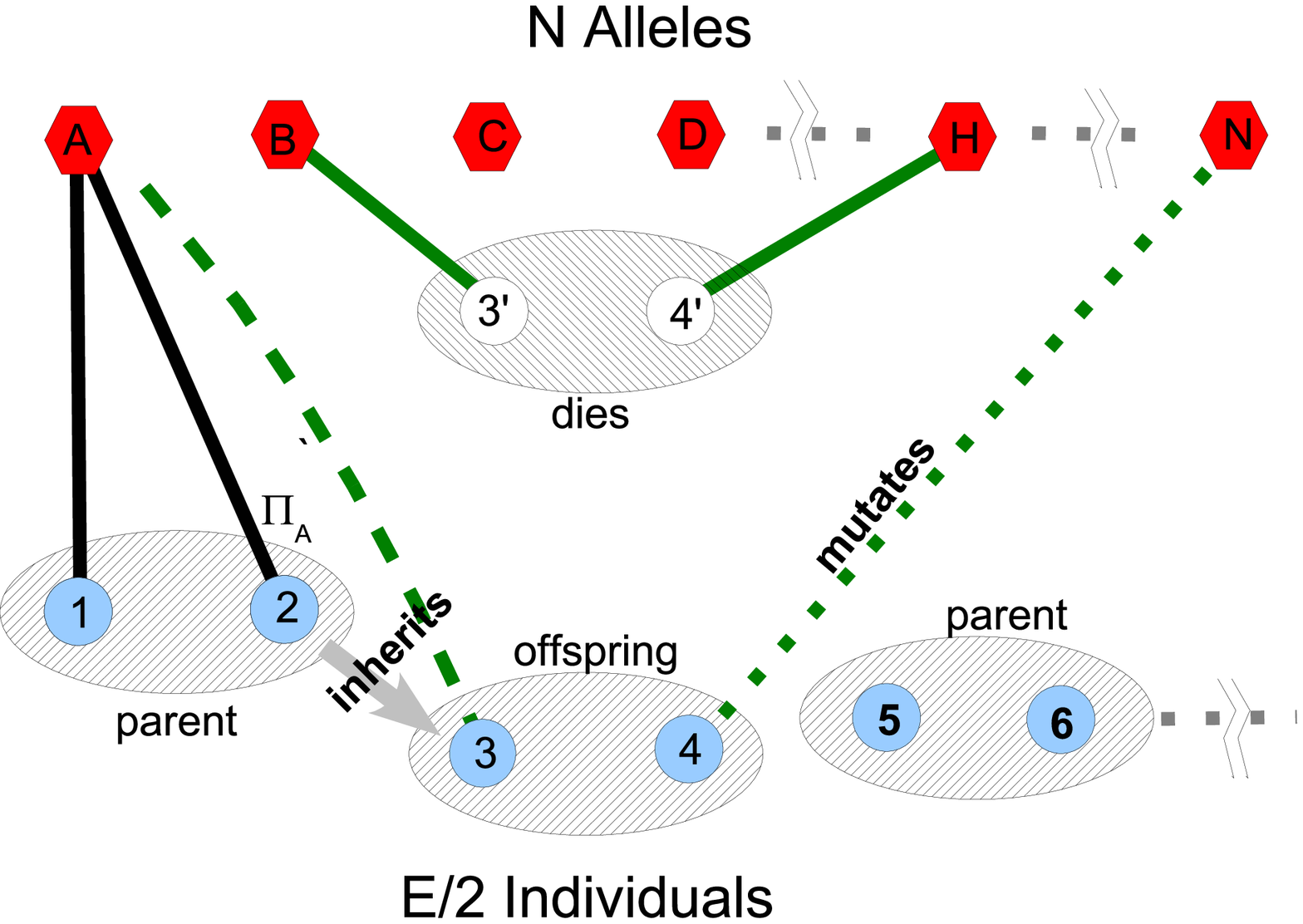}}}
\caption{Interpretation of our model as a diploid gene inheritance
and mutation model. Organism (3',4') dies and is replaced by (3,4)
whose parents are the organisms (1,2) and (5,6).  From the (1,2)
parent it inherits copy number 2 of the gene which is allele A.
However the gene it inherits from the (5,6) parent mutates to allele
N. }
 \label{fdiploid}
\end{figure}

In a diploid cell\tnote{See section 3.2 of \cite{BG99}.} there are
two copies of a gene and most cells of most higher organisms are
of this type.  Ignoring fitness etc.\ we can see that we can
represent the allele frequencies of one gene in a constant
population of diploid cells with the usual rewiring model  as
shown in Fig.~\ref{fdiploid}.\tnote{There pairs of individuals
represent an organism, each individual vertex is one copy of the
gene in that organism so its linked by an edge the one alleles, an
artifact vertex.  Note that notation may vary as the number of
organisms is $E/2$ in the notation of this paper. Also there are
two gene copies per time step in most diploid models while in the
notation here we have one per time step.  Thus the mutation rate
of a diploid model $u$ is going to be $p_r/2$ in our language. I
thought that the probability that an organism has no mutation is
roughly (1-2u) so $p_r = 2u$ and another factor of two appears.}

There is also a close relationship between our network model and
various models of statistical physics, a connection already noted in
some places \cite{EH05,OYT05}. In the original Urn model of the
Ehrenfests \cite{EE07} one has $E$ balls placed in two Urns. At
random times given by a Poisson process, a ball chosen at random is
moved from one urn to another. This corresponds to a continuous time
version of our model with the artifacts being the Urns so $N=2$ and
the individuals represent the balls.  Choosing $p_r=1$ in our model
reproduces the behaviour of the original Urn model.

There is one subtlety in that in the original Urn model the ball is
never put back into the Urn it was drawn from.  These events are
allowed in our model and are precisely the ones which require the
factors of $(1-\Pi)$ in our master equation \tref{neqngen} as they
leave the configuration unchanged.  The difference between the
original Urn model and our model for $N=2$, $p_r=1$ and continuous
time is just a matter of a factor of two in the rates.

However generic Urn models are often encountered in some obvious
variations of the Ehrenfest version, in particular with $N$ urns and
with different forms for the rate at which balls are moved and where
they are then placed \cite{GBM95,GL02,OYT05,OYT06}. Some of these
variations of the original Urn model are equivalent to other models
such as the Backgammon or Balls-In-Box models used for glasses
\cite{Ritort95,BBJ97}, simplicial gravity \cite{BBJ99} and wealth
distributions \cite{BDJKNPZ02}.\tnote{Now the individual vertices
represent the balls while the artifact vertices are the boxes into
which the balls are placed, as indicated by the edges. The
difference is that each box (artifact $a$) is assigned an energy
$\mathcal{E}(k_a)$ depending on the number of balls $k_a$ in its box
alone.  This may be thought of as an undressed contribution to the
overall probability for a given configuration $\Pi_{a=1}^N \exp
\{-\mathcal{E}(k_a)\}$. The key is that this is a micro-canonical
ensemble and the total number of balls is fixed to be $N$.  Thus the
balls fall into the boxes with dressed probabilities $p(k)$ which
have to be determined in the Balls-in-Box approach but which
correspond to the input artifact probability distribution in our
rewiring model.}  The zero range processes
\cite{EvansMR00,EH05,PM05} can also be interpreted as Urn models,
with the `misanthrope' process on a fully connected geometry being
closest to our basic model.

Using the terminology of the Urn model review \cite{GL02}, the
`geometry' of the Urn model refers to which Urns are connected ---
an artifact network in our model as discussed in Section
\ref{ssgenmodel}.  The simplest `mean-field' geometry, i.e.\ a
complete graph for the artifact network, is what we assume in our
simple model.  On the other hand the basic zero-range process models
\cite{EvansMR00,EH05} use a one-dimensional ring. If we allow
processes where the ball is placed back into the urn it came from,
then the rate at which a ball moves is given by $u(d,a)$ per ball
where $d$ is the departure urn and $a$ the arrival urn. Usually the
rates used factorise into two terms, one depending only on the
number of balls in the departure urn $k_d$ (number of edges, i.e.\
the artifact vertex degree) and the other on the number of balls in
the arrival urn $k_a$.  In our terminology $u(d,a) = \Pi_R(k_d)
\Pi_A(k_a)$. Then the three rules for ball selection discussed in
\cite{GBM95,GL02} correspond to our generalisation \tref{PiRPiAfull}
as follows: Rule A (random ball to random urn, Ehrenfest class) is
our $q_p=1$, $p_r=1$; Rule B (random urn to random urn, Monkey
class) is our $q_a=1$, $p_r=1$; Rule C (random ball to random ball)
is our $q_p=1$, $p_p=1$.

However we stress that to include processes where balls are returned
to the urn they were drawn from, the master equation has to contain
the factors $(1-\Pi)$ while such terms are normally absent in the
evolution equations of literature on Urn and related models e.g.\ in
\cite{GBM95,GL02,OYT05}.  If we were to exclude such events then our
transition rates $u(d,a)$ will not factorise into departure and
arrival dependent terms\footnote{One exception is for a two Urn
model as then the number of balls in the arrival urn may always be
written in terms of those in the departure Urn and vice versa. This
means typical expressions for rates are always factorisable.}. For
instance in our language the factor normally associated with the
arrival vertex, our attachment probabilities $\Pi_A$, will have to
depend on the departure urn as well as on the properties of the
arrival urn, and for us would take the form \tref{PiAnotad} while
the literature usually uses simple factorisable forms e.g.
\cite{GBM95,GL02,OYT05,OYT06}. As we noted in Section
\ref{ssgenmodel} this will be important when a significant fraction
of balls are in any one box, as is the case with a condensate.

There is a way round this problem and that is to work with our
solution in continuous time \tref{conttime} and then to rescale our
time $t$ back into the time $t_\mathrm{urn}$ of an Urn model where
one can not put the ball back into the urn it was drawn from. From
the number of these events allowed in our model but excluded from an
Urn model we have for infinitesimal time steps
\beq
 \mathrm{d}t-\mathrm{d}t_\mathrm{urn}
  = \left[ \frac{p_r}{N} + \frac{p_p \langle k^2 \rangle}{E \kav} \right]\mathrm{d}t
\eeq
where the second moment $\langle k^2 \rangle$ is easily derived from
$F_2(t)$ of \tref{eqF2tres}.

Finally we note that many models in sociophysics may be cast as
generalisations of our bipartite rewiring model.  If we add an
Individual graph then for a copying (preferential attachment) event,
an individual copies the artifact choice made by one of its
neighbours in the Individual graph. When $p_p=1$ and $N=2$ this is
the basic Voter model \cite{Liggett85}, as used for instance for
language evolution \cite{SCEM06}.\tnote{What about $\rho$ of the
language papers?}  Our results are equivalent to having an
Individual graph which is complete with tadpoles\footnote{If this is
performed on just a complete graph then this is an $N=2$ Urn model
with rule C of \cite{GBM95,GL02} performed on mean field Urn
geometry.}. Our time scale is $(\tau_2/E) = E/2$ which agrees with
the $O(E)$ result quoted in \cite{SR05}. However our result shows
the effect on both the consensus and on the time scale to reach
equilibrium of adding some randomness to such Voter models.

Our results may also be useful in other sociophysics models. In one
variation of the Minority Game \cite{ATBK04} individuals choose the
`best' strategy known to them, comparing their own against all those
used by their neighbours as defined by an Individual graph. Each
artifact vertex in our model would then represent a different
strategy. In this case what is best is continually changing as
generally the more popular one strategy becomes the less successful
it will be. Thus statistically, it is likely that the resulting
instantaneous artifact degree distribution $n(k,t)$ will be
indistinguishable from that obtained by just copying the artifact of
a random neighbour which as a simple random walk is likely to lead
to effective preferential attachment. It is no surprise then that
the long time results for the popularity of strategies in
\cite{ATBK04} follows a simple inverse power law with a large degree
cutoff, the form found in \tref{ngammadist}.

\section{Summary and Conclusions}\label{sec:sumconcl}

The starting point of our work is the observation that the usual
mean field master equations seen for network evolution are not
suitable for general rewiring problems.  One needs to add the
factors of $(1-\Pi)$ seen in \tref{neqngen} if the degree
distribution is to behave properly at the maximum degree
\cite{Evans06}. With these terms and the simplest case of linear
attachment/removal probabilities the \emph{exact} solution for the
degree distribution at \emph{any} time can be found for
\emph{arbitrary} values of the parameters, here expressed in terms
of the generating function $G(z,t)$ of \tref{Gktdef}, \tref{Gkmdef}
and \tref{Gmresult}. This is better than can be done for simple
growing networks where the exact equilibrium solution is known for
simple attachment probabilities but the finite size (finite time)
system corrections are only known asymptotically (e.g.\ see
\cite{DM01,KR01,ES05}). Previous results for equivalent models give
results that are only approximations, often for infinitely large
systems in equilibrium though all are consistent with the results
derived here\footnote{The literature does however tackle other more
complicated variations of the model not considered here.}.  We have
also compared our analytic results against numerical simulation in
several ways and seen that agreement is excellent. We know of no
other network model that has the exact time dependent solution for
arbitrary parameters range and suggest that this model may prove to
be as useful a model as the Erd\H{o}s-R\'eyni random graph has
been.\tnote{Can we find examples of calculations where degree
distribution only is needed?}

In particular the equilibrium degree distribution of \cite{Evans06}
is found as the long time solution.  It has two characteristic
regimes: if when all edges have been rewired once (on average) at
least one rewiring was done randomly then a simple inverse power law
with exponential cutoff is obtained, otherwise we have a regime with
a condensate.

We have confirmed the slow approach to equilibrium and the
conjectured form for the second eigenvalue $\lambda_1$ of
\cite{EP06ECCS}. However here we have shown that the long time
evolution is governed not by the second largest eigenvalue but the
third largest, $\lambda_2 = 1 - (2p_r/E) - (2p_p/E^2)$ with
associated time scale $\tau_2 = -1/\ln(\lambda_2)$.

We have also noted that this simple bipartite graph rewiring model
captures the degree distribution of many other networks, with that
of the original Watts and Strogatz model \cite{WS98} as one limit of
our model.  In particular we have the exact degree distribution at
any time and any parameter value for the rewiring of a general
random graph.  From this we can obtain various global properties
analytically as a function of time using various known formulae
\cite{MR95,MR98,NSW01,DMS03a}. However many of the alternative
realisations require no explicit network as in the link to
Urn/Backgammon/Balls-in-Boxes models and zero range processes
\cite{OYT05,OYT06,BBBJ00,Ritort95,BBJ97,BBJ99,BDJKNPZ02,EvansMR00,EH05,PM05}.

The model also has a wide range of practical applications.  As most
practical systems can not grow indefinitely, this fixed sized
rewiring model will often be more appropriate than a growing network
model. The Urn-type models have been applied to glasses
\cite{Ritort95}, simplicical gravity \cite{BBJ99} and wealth
distributions \cite{BDJKNPZ02}.  Models for social science in both
modern and archaeological contexts
\cite{Neiman95,BS03,HB03,BHS04,HBH04} can be be cast as our model.
The applications in these papers include baby name frequencies
\cite{HB03}, pedigree dog breed popularity \cite{HBH04} and pottery
styles \cite{Neiman95,BS03}.  Sociophysics models may also be
related to our work.  The Voter model \cite{Liggett85,SR05}, as
applied to language evolution \cite{SCEM06}, and the choice of
strategy in a Minority game variant \cite{ATBK04}, may be linked to
our bipartite graph approach. Finally basic models of population
genetics \cite{KC64,CK70} and more generally any process where
inheritance is important, such as with Family names \cite{ZM01}, can
be encoded by our network rewiring.

Many of the related examples in the literature also study cases
beyond our simple model, for example non-linear attachment
probabilities $\Pi_A \propto k^\beta$ for $\beta \in \mathbb{R}$,
attachment probabilities whose scale varies with the artifact
(\emph{fitness}), pure Artifact or pure Individual graphs, and
growing systems $(dE/dt) \neq 0$. These can often be captured by
extensions to our basic model but the downside of this
sophistication is that the mean field equations are then only
approximations whose exact algebraic solution is probably
unobtainable in any case. Rather the literature usually works in a
large network long time approximation, $E \gg 1$, often in a
particular part of parameter space such as $1 \gg p_r \gg E^{-1}$ or
$p_r=0$. We though have exploited the simplicity of our model in
order to obtain exact solutions for any time or parameter value.

At worst this simple bipartite model provides a useful null model
against which to test other hypotheses \cite{BS05}. However we have
also argued in Section \ref{ssgenmodel} why copying may be a more
widespread method than the obvious cases involving inheritance
mechanisms.

Finally we note the scaling properties of the model. In many
practical examples the artifacts are really categories imposed by
investigators on a collection of objects.  In almost all cases, each
object could be individually identified if one wishes. Indeed the
objects may be being chosen by individuals based on characteristics
completely different from those recorded by the researcher.  No
pedigree dog \cite{HBH04} is genetically pure, a personal
\cite{HB03} or family name \cite{ZM01} may come in several close
variations, and who assigns a particular style to an archaeological
pottery find \cite{Neiman95,BS03}?

Consider an exemplary small study \cite{MS06} where the shoes of
around two hundred male physics students leaving a lecture were
photographed. Various researchers categorised them in completely
different ways giving different degree distributions from the same
data. For instance one could categorise each shoe by colour,
material and fastening method. Still what constitutes say a `blue'
shoe may be a context dependent matter of physical and social
perception so researchers and wearers may not even agree how to
classify a given shoe under the one scheme.

So if such artifact popularity distributions are to have much
meaning they ought to be largely independent of this categorisation.
Thus consider pairing the artifacts at random and calculating the
degree distribution for these pairs, $n_2(k)$, even though the model
continues to make its rewiring selections based on the original
single artifact vertices. That is at each event we choose to make a
preferential (copying, inheritance) attachment or a random
(innovation, mutation) attachment to the artifact pairs with exactly
the same probability $p_p$ and $p_r$. Given our \emph{linear}
attachment/removal probabilities the effective probability for
attaching to a given artifact pair is just the sum of the degrees of
its constituent artifacts, i.e.\ it is still proportional to the
degree of the artifact pair. On the other hand the probability we
attach to a given artifact chosen at random is halved but only
because the number of artifact pairs $N_2$ is just half the original
number of artifacts, $N\rightarrow N_2= N/2$. With the number of
edges $E$ unchanged, changing $N$ to $N_2$ is the only change we
need to make in our equations. Vitally, the form for the attachment
and removal probabilities remains the same and thus the form of the
solutions is unchanged.  We will get the same qualitative behaviour,
a power law with an exponential cutoff.  In such cases the cutoff
$\zeta$ \tref{zetasol} remains unchanged and only the slope $\gamma$
\tref{gammasimp} changes in comparing $n(k)$ to the pair degree
distribution $n_2(k)$. However the slope is invariably
indistinguishable from one in a practical data set or it will be
unmeasurable with a small cutoff $\zeta$. Thus for a linear
attachment plus random attachment model, the distribution of
artifact choice is independent of how artifacts are classified for
all practical purposes.\tnote{Variation with random attachment to
active vertices. Note on how its the large $k$ behaviour of
$\Pi_A/\Pi_R$ that decides condensation.}

\section*{Acknowledgements}

TSE thanks H.Morgan and W.Swanell for useful conversations.

\tcomment{
\appendix

\section{Additional Material}

This material will not be in the published version.  It is given for
convenience.

The Hypergeometric function can be defined as the series
\beq
 F(a,b;c;z) = \sum_{n=0}
 \frac{\Gamma(a+n)}{\Gamma(a)}
 \frac{\Gamma(b+n)}{\Gamma(b)}
 \frac{\Gamma(c)}{\Gamma(c+n)} \frac{1}{n!} z^n
\eeq
The sum is terminated at $-a$ if $a$ is a negative integer, likewise
for $b$.  The Pochhammer symbols are the ratios
\beq
 (a)_n := \frac{\Gamma(a+n)}{\Gamma(a)}
\eeq
and are often useful.

The Hypergeometric differential equation is\tnote{See \cite{GR}
9.151, 9.152.}
\beq
z(1-z) f^{\prime\prime} + [c-(a+b+1)z]f^\prime - abf =0
\eeq
has two linearly independent solutions solution.  For $c$
non-integer the solution is
\beq
 f(z) = A F(a,b;c;z) + Bz^{1-c} F(a-c+1,b-c+1;2-c;z)
\eeq
The differential is given by\tnote{From \cite{Erd}, p.102, eqn.\
20.}
\beq
\frac{d^n}{dz^n}F(a,b;c;z) =
 \frac{\Gamma(a+n)}{\Gamma(a)}
 \frac{\Gamma(b+n)}{\Gamma(b)}
 \frac{\Gamma(c)}{\Gamma(c+n)}
 F(a+n,b+n;c+n;z)
\eeq
and a useful special value is \tnote{\cite{Erd} p.104, (46).}
\beq
 F(a,b;c;1) =
 \frac{\Gamma(c) \Gamma(c-a-b)}
 {\Gamma(c-a) \Gamma(c-b)}
\eeq

The asymptotic limit of a $\Gamma$ function is
\beq
 \ln \left( \Gamma (z) \right) = z \ln (z) -z - \half \ln(z) + \ln
 (\sqrt{2 \pi}) + O(1/z)
 \label{GammaAsymp}
\eeq
so\tnote{From \cite{GR} [8.328.2].}
\beq
 \lim_{|z| \rightarrow \infty} \frac{\Gamma(z+a)}{\Gamma(z)}
  = \exp \{  a \ln(z)   \} = z^a
\eeq
and
\beq
 \frac{\Gamma(1-x)}{\Gamma(1-x-n)}
 =
 (-1)^n \frac{\Gamma(x+n)}{\Gamma(x)}
\eeq

Finally for expansions it is useful to have
\bea
\Gamma(z+\epsilon) &=& \Gamma ( z)
 \left( 1+ \epsilon \psi(z) \right) ,
 \; \; \; z \not\in \{0, \mathbb{Z}^- \} \label{Gamnp}
 \\
 \Gamma(-n+\epsilon) &=& \frac{1}{\epsilon}
\frac{(-1)^{n}}{\Gamma(1+n)} \left( 1+ \epsilon \psi(1+n) \right) ,
\; \; \; n \in \{ 0, \mathbb{Z}^+ \} \label{Gamnn}
\\
 \psi(z) &:=&
 \frac{d \ln( \Gamma ( z) ) } {dz}
 \label{psidef}
 \\
 \psi(1+x) &=& \psi(x) + \frac{1}{x}
 \label{psix}
 \\
 \psi(1) &=& -\gamma \approx - 0.577 215 664 90 , \; \; \; \gamma
 \mbox{ is Euler's constant},
 \label{psi1}
 \end{eqnarray}

}



\begin{thebibliography}{99}


\bibitem{WS98}
 D.J.Watts and S.H.Strogatz,
 \tarttitle{Collective dynamics of `small-world' networks}
 Nature \vol{393} 440 (1998).

\bibitem{BCK01}
  Z.Burda, J.D.Correia and A.Krzywicki,
  \tarttitle{Statistical ensemble of scale-free random graphs}
  Phys.Rev.E\vol{64} 046118 (2001).

\bibitem{DM03}
  S.N. Dorogovtsev and J.F.F.Mendes,
  \tbktitle{Evolution of Networks: From Biological Nets to the Internet
 and WWW}
 (Oxford University Press, Oxford, 2003).

\bibitem{DMS03}
  S.N. Dorogovtsev, J.F.F.Mendes and A.Samukhin,
  \tarttitle{Principles of statistical mechanics of uncorrelated random networks}
  Nucl.Phys.B\vol{666} 396 (2003)




\bibitem{PLY05}
 K.Park, Y.-C.Lai and N.Ye,
 \tarttitle{Self-organized scale-free networks}
 Phys.Rev.E\vol{72} 026131 (2005).

\bibitem{XZW05}
 Y.-B. Xie, T.Zhou and B-.H.Wang,
 \tpretitle{Scale-free networks without growth}
 Report \eprint{cond-mat/0512485}.

\bibitem{OTH05}
 J.Ohkubo, K.Tanaka and T.Horiguchi,
 \tpretitle{Generation of complex bipartite graphs using a preferential rewiring process}
 Phys.Rev.E\vol{72} 036120 (2005).

\bibitem{GBM95}
 C.Godr\`{e}che, J.P. Bouchaud and M.M\'{e}zard,
  \tarttitle{Entropy barriers and slow relaxation in some random walk models}
  J.Phys.A \vol{28} L603 (1995).


\bibitem{GL02}
  C.Godr\`{e}che and J.~M.Luck,
  \tarttitle{Nonequilibrium dynamics of urn models}
  J.Phys.Cond.Matter \vol{14} 1601 (2002).

\bibitem{OYT05}
 J.Ohkubo, M.Yasuda and K.Tanaka,
 \tpretitle{Preferential urn model and non-growing networks}
 Phys.Rev.E\vol{72} 065104(R) (2005).

\bibitem{OYT06}
 J.Ohkubo, M.Yasuda and K.Tanaka,
 \tpretitle{Replica analysis of a preferential urn model}
 Report \eprint{cond-mat/0601193}.


\bibitem{BBBJ00}
    P.Bialas, L.Bogacz, Z.Burda and D.Johnston
    \tarttitle{Finite size scaling of the balls in boxes model}
    Nucl.Phys. B575 (2000) 599-612.


\bibitem{Ritort95}
  F.Ritort,
  \tarttitle{Glassiness in a model without energy barriers}
  Phys.Rev.Lett.\ \vol{75} 1190  (1995).


\bibitem{BBJ97}
    P.Bialas, Z.Burda and D.Johnston,
    \tarttitle{Condensation in the Backgammon model}
    Nucl.Phys.B\vol{493} 505  (1997).

\bibitem{BBJ99}
    P.Bialas, Z.Burda and D.Johnston,
    \tarttitle{Phase diagram of the mean field model of simplicial gravity}
    Nucl.Phys. B542 413  (1999).

\bibitem{BDJKNPZ02}
    Z.Burda, D.Johnston, J.Jurkiewicz, M.Kaminski, M.A.Nowak, G.Papp and
    I.Zahed,
    \tarttitle{Wealth Condensation in Pareto Macro-Economies}
    Phys.Rev.E\vol{65} 026102  (2002)

\bibitem{EvansMR00}
  M.R.Evans,
 \tarttitle{Phase transitions in one-dimensional nonequilibrium systems}
 Braz.J.Phys.\ \vol{30} 42 (2000).

\bibitem{EH05}
  M.R.Evans and  T.Hanney,
 \tarttitle{Nonequilibrium statistical mechanics of the zero-range process
and related models}
 J.Phys.A. \vol{38} R195-R240 (2005).

\bibitem{PM05}
 O.Pulkkinen and J.Merikoski,
 \tarttitle{Phase transitions on Markovian bipartite graphs: an application of the zero-range process}
 J.Stat.Phys.\ \vol{119} 881 (2005).


\bibitem{Neiman95}
 F.D.Neiman,
 \tpaptitle{Stylistic variation in evolutionary perspective:
Inferences from decorative diversity and inter-assemblage distance
in Illinois Woodland Ceramic assemblages}
 American Antiquity \vol{60} 1  (1995).

\bibitem{BS03}
 R.A.Bentley and S.J.Shennan,
  \tarttitle{Cultural Transmission and Stochastic Network Growth}
  American Antiquity \vol{68} 459 (2003).


\bibitem{HB03}
 M.W.Hahn and R.A.Bentley,
  \tarttitle{Drift as a Mechanism for Cultural Change: an example from baby names}
  Proc.R.Soc.Lon.B\vol{270} S120 (2003).

\bibitem{HBH04}
 H.A.Herzog, R.A.Bentley and M.W.Hahn,
 \tpaptitle{Random drift and large shifts in popularity of dog
 breeds}
 Proc.R.Soc.Lon B (Suppl.) \vol{271} S353  (2004).

\bibitem{BHS04}
 R.A.Bentley, M.W.Hahn and S.J.Shennan,
  \tarttitle{Random Drift and Cultural Change}
  Proc.R.Soc.Lon.B\vol{271} 1443 (2004).

\bibitem{BS05}
 R.A.Bentley and S.J.Shennan,
  \tarttitle{Random Copying and Cultural Evolution}
  Science \vol{309} 877 (2005).

\bibitem{ZM01}
 D.Zanette and S.Manrubia,
 \tarttitle{Vertical transmission of culture and the distribiution
 of family names}
 Physica A \vol{295} 1 (2001).

\bibitem{KC64}
 M.Kimura and J.F.Crow,
 \tpaptitle{The Number of Alleles that can be Maintained in a Finite
 Population}
 Genetics \vol{49} 725  (1964).

\bibitem{CK70}
 J.F.Crow and M.Kimura,
 \tbktitle{An Introduction to Population Genetics Theory}
 (Harper and Row, New York, 1970).

\bibitem{Liggett85}
 T.M.Liggett,
 \tbktitle{Interacting Particle Systems}
 (Springer-Verlag, New York, 1985).

\bibitem{SR05}
  V.Sood and S.Redner,
  \tarttitle{Voter Model on Heterogeneous Graphs}
  Phys.Rev.Lett. \vol{94} 178701 (2005).

\bibitem{SCEM06}
  D.Stauffer, X.Castello, V.M.Eguiluz and M.San Miguel,
  \tarttitle{Microscopic Abrams-Strogatz model of language competition}
  Physica A \vol{374} 835 (2006).

\bibitem{ATBK04}
 M.Anghel, Z.Toroczkai, K.E.Bassler and G.Korniss,
  \tpaptitle{Competition in Social Networks: Emergence of a Scale-Free Leadership Structure
  and Collective Efficiency}
  Phys.Rev.Lett \vol{92} 058701  (2004).

\bibitem{DM01}
  S.N.Dorogovtsev and J.F.F.Mendes,
 \tpaptitle{Evolution of networks}
 Adv.Phys.\ \vol{51} 1079  (2002).

\bibitem{KR01} P.L.Krapivsky and S.Redner,
 \tarttitle{Organisation of Growing Random Networks}
 Phys.Rev.E \vol{63} 066123  (2001).

\bibitem{KR02} P.L.Krapivsky and S.Redner,
 \tarttitle{Finiteness and Fluctuations in Growing Networks}
 J.Phys.A \vol{35} 9517  (2002).

\bibitem{ES05}
 T.S.Evans and J.P.Saram\"aki,
 \tarttitle{Scale Free Networks from Self-Organisation}
 Phys.Rev.E \vol{72}  026138  (2005).


\bibitem{Evans06}
 T.S.Evans,
  \tpretitle{Exact Solutions for Network Rewiring Models}
  Eur.\ Phys. J. B \vol{56} (2007) 65-69
 [\eprint{cond-mat/0607196}].

\bibitem{EP06ECCS}
 T.S.Evans and A.D.K.Plato,
  \tpretitle{Exact Solutions for Models of Cultural Transmission and Network Rewiring}
  (to appear in the proceedings of ECCS06)
  Report \eprint{physics/0608052}.

\bibitem{MR95}
  M.Molloy and B.Reed,
  \tarttitle{A critical point for random graphs with a given degree sequence}
  Random Structures \& Algorithms, \vol{6} 161 (1995).

 \bibitem{KRL00}
 P.L.Krapivsky, S.Redner and F.Leyvraz,
 \tpretitle{Connectivity of Growing Random Networks}
 Phys.Rev.Lett. \vol{85} 4629  (2000).

\bibitem{Vaz00}
  A.V\'azquez,
  \tpretitle{Knowing a network by walking on it: emergence of scaling}
  Report \eprint{cond-mat/0006132}.

\bibitem{Vaz02}
  A.V\'azquez,
  \tpaptitle{Growing networks with local rules:
  preferential attachment, clustering hierarchy and degree correlations}
  Phys. Rev.E \vol{67}  056104  (2003).


\bibitem{SK04}
 J.Saram\"aki and K.Kaski,
 \tarttitle{Scale-Free Networks Generated by Random Walkers}
 Physica A\vol{341} 80 (2004).


\bibitem{MR98}
  M.Molloy and B.Reed,
  \tarttitle{The size of the giant component of a random graph with a given degree sequence}
  Combin.Probab.Comput. \vol{7} (1998) 295

\bibitem{NSW01}
  M.E.J.Newman, S.H.Strogatz and D.J.Watts,
  \tarttitle{Random graphs with arbitrary degree distributions and their applications}
  Phys.Rev.E \vol{64} 026118 2001.

\bibitem{DMS03a}
    S.N.Dorogovtsev, J.F.F.Mendes and A.Samukhin,
    \tarttitle{Metric structure of random networks}
  Nucl.Phys.B \vol{653} 307 (2003).

\bibitem{EE07}
  P.Ehrenfest and T.Ehrenfest,
  Physik Z.\ \vol{8} 311 (1907).

\bibitem{MS06}
 H.Morgan and W.Swanell,
 Imperial College London BSc project reports (2006).



\end{thebibliography}
\end{document}